\documentclass[acmsmall, manuscript,screen]{acmart}
\usepackage{graphicx}
\AtBeginDocument{%
  }

\usepackage{subcaption}
\usepackage{multirow}       
\usepackage{adjustbox}
\usepackage{xcolor,colortbl}
\usepackage{color, colortbl} 
\usepackage{nicematrix}

\newcommand{\titleFullNameFirst}{Call of Duty\textregistered: Modern Warfare\textregistered II}
\newcommand{\titleFullName}{Call of Duty: Modern Warfare II}
\newcommand{\titleShort}{COD:MWII}

\newcommand{\company}{ABK}

\newcommand{\modMILD}{Milder}
\newcommand{\modSEVERE}{Stricter}

\definecolor{highlight}{rgb}{1.0,0.90,0.8}	
\definecolor{Gray}{gray}{0.96}


\begin{document}

\title{Online Moderation in Competitive Action Games: How Intervention Affects Player Behaviors}

\author{Zhuofang Li}
\authornote{Both authors contributed equally to this research.}
\email{zhuofang@caltech.edu}
\author{Rafal Kocielnik}
\authornotemark[1]
\email{rafalko@caltech.edu}
\author{Mitchell Linegar}
\email{mlinegar@caltech.edu}
\author{Deshawn Sambrano}
\email{sambrano@caltech.edu}
\affiliation{%
  \institution{California Institute of Technology}
  \city{Pasadena}
  \country{USA}
}

\author{Fereshteh Soltani}
\email{feri.soltani@activision.com}
\author{Min Kim}
\email{min.kim@activision.com}
\author{Nabiha Naqvie}
\email{nabiha.naqvie@activision.com}
\author{Grant Cahill}
\email{grant.cahill@activision.com}
\affiliation{%
  \institution{Activision-Blizzard-King}
  \city{Los Angeles}
  \country{USA}
}

\author{Animashree Anandkumar}
\email{anima@caltech.edu}
\author{R. Michael Alvarez}
\email{rma@caltech.edu}
\affiliation{%
  \institution{California Institute of Technology}
  \city{Pasadena}
  \country{USA}
}


\renewcommand{\shortauthors}{Li and Kocielnik, et al.}

\begin{abstract}
  Online competitive action games have flourished as a space for entertainment and social connections, yet they face challenges from a small percentage of players engaging in disruptive behaviors. This study delves into the under-explored realm of understanding the effects of moderation on player behavior within online gaming on an example of a popular title - \titleFullNameFirst{}. 
  We employ a quasi-experimental design and causal inference techniques to examine the impact of moderation in a real-world industry-scale moderation system. We further delve into novel aspects around the impact of delayed moderation, as well as the severity of applied punishment. We examine these effects on a set of four disruptive behaviors including cheating, offensive user name, chat, and voice. 
  Our findings uncover the dual impact moderation has on reducing disruptive behavior and discouraging disruptive players from participating. We further uncover differences in the effectiveness of quick and delayed moderation and the varying severity of punishment. Our examination of real-world gaming interactions sets a precedent in understanding the effectiveness of moderation and its impact on player behavior.
  Our insights offer actionable suggestions for the most promising avenues for improving real-world moderation practices, as well as the heterogeneous impact moderation has on indifferent players.
  
\end{abstract}

\maketitle

\section{Introduction}

The proliferation of online social interactions and competitive action games has enriched the virtual landscape, providing opportunities for entertainment, improved well-being, and social connections \cite{kriz2020gaming, bourgonjon2016players}. However, this digital frontier is not without challenges. While the majority of players engage in respectful and enjoyable gameplay, a small percentage have leveraged these platforms to exhibit disruptive behaviors such as cheating, trolling, and offensive speech \cite{cook2019whom}. Moderation, or the regulation of user behavior by platforms, has thus emerged as an essential component of online gaming. The process demands tangible resources, including employees, infrastructure, and time, particularly when human review is required to mitigate complex challenges that can't easily be handled via automation alone, such as domain shift  (changes in expression of disruptive behavior or its definition) \cite{srikanth2021dynamic} and strategic classification (when the players strategically alter their behavior to circumvent automated systems and avoid detection) \cite{frommel2022effective}. 

Platforms face tangible constraints in moderating user behavior. Though algorithmic tools can lessen the content requiring review, the human element is vital \cite{Contentm60:online}. This is due to aspects challenging to automate such as interpretation ambiguity \cite{beres2021don}, the need for broader contextual understanding \cite{frommel2022effective}, and the need for common-sense judgment. Optimal moderation efforts often require collaboration between human judgment and technological tools \cite{rieder2021fabrics, link2016human}. Unfortunately, the ratio of human moderators to the volume of content requiring moderation leads to bottlenecks in the review process \cite{gorwa2020algorithmic}. Thus many important scientific challenges exist for moderating distributive player behavior \cite{kocielnik6challenges}.

\paragraph{\textbf{Prior work:}} Despite the importance of moderation, empirical work exploring the causal effects of this practice within the gaming community remains sparse \cite{wijkstra2023help}, with most work focusing more on developing novel data-driven approaches to detecting toxic behaviors \cite{canossa2021honor, weld2021conda}, testing theory-informed hypotheses related to the emergence of toxicity \cite{kwak2015exploring}, or studying the toxicity in various social communication platforms rather than in games themselves \cite{ghosh2021analyzing}. 
These studies, however, don't examine the effects that moderation of toxic behavior has on players in real-world gaming situations. The few prior studies that do examine the impact of moderation relied only on small sample survey-based examinations around the self-reported perceptions of players \cite{ma2023transparency, kou2021flag, kordyaka2021curing} or moderators \cite{cullen2022practicing, aguerri2023enemy}. Both of these lack the scale to draw conclusions about the effectiveness of different types and properties of moderation at scale in real-world gaming titles. Indeed a recent review of intervention systems for toxicity highlighted that only a few interventions are evaluated with players and in commercial settings \cite{wijkstra2023help}, highlighting the potential for more research with higher external validity. 
Our study fills this gap by examining real-world large-scale moderation data from one of the more renowned titles in the industry - 2022's \emph{\titleFullName{} (\titleShort)}.

\paragraph{\textbf{Our work:}} In this paper, we utilize a quasi-experimental design and the latest causal machine-learning methods (causalML)\cite{kaddour2022causal} to analyze the impact of moderation on player behavior. We specifically examine and compare the behavior of players who were moderated as compared to those who were not in terms of the impact on offensive behavior (repeated offenses) and the number of days with matches played (participation rate). We focus on players who were eventually subjected to human moderation (to control for false reports) and control for consistent types of behavior. Taking into account the principle of immediacy, we also examine the effect of delayed consequences by examining the player's behavioral measures post-moderation in the context of delayed versus quick interventions. Finally, we evaluate the impact the severity of applied moderation actions has on immediate post-moderation player behavior.

\paragraph{\textbf{Findings:}} Our results reveal a dual impact of moderation on reducing disruptive behavior and on discouraging some disruptive players from participating. We also uncover trade-offs between quick and delayed moderation, as well as varying severity of punishment. Specifically, our analysis of player behavior shows that moderation effectively lowers disruptive behavior by up to 70\% but can lead to up to 12\% fewer matches played per day by disruptive players. Quick moderation is more effective at reducing disruptive behavior. Cheaters seem to respond differently to moderation, in that moderation results in a larger reduction of participation (days with matches) for these players than for toxic players. These results offer insights into how moderation affects participation rate and disruptive behavior, including the unique response of cheaters.

\paragraph{\textbf{Contributions:}} This study offers several significant contributions:

\begin{enumerate}
    \item We present one of the first studies examining large-scale real-world moderation efforts from one of the most popular gaming titles - \titleFullName{} (\titleShort).
    
    \item We uncover an important dual impact of moderation in terms of reducing disruptive behavior and discouraging disruptive players from participation, as well as the importance and impact of quick versus delayed moderation.
    
    \item Our findings lead to actionable insight into the moderation practices, such as the need for quick moderation and the need to understand why some disruptive players change their behavior, while others reduce their participation. We provide a discussion of the real-world implications, setting the agenda for comprehensive analysis and future opportunities.
    \item Our study highlights effective strategies for mitigating disruptive behaviors in online environments, offering guidance for gaming and online platforms to foster safer, more engaging communities. By examining how different moderation techniques influence player behavior, our findings provide actionable knowledge that extends beyond gaming to inform digital interaction and community management across various platforms.

\end{enumerate}

\section{Background and Related Work}

\subsection{Online Gaming and the Challenge of Toxicity}

Online multiplayer video games, such as \titleShort, have become a significant part of contemporary entertainment. These games offer various benefits, including satisfaction of psychological needs, facilitation of social relationships, and aid in coping mechanisms \cite{kriz2020gaming, raith2021massively, trepte2012social}. The real-time interaction in these games enhances user experience through increased enjoyment and social exchange \cite{bourgonjon2016players, kordyaka2021curing}.

However, the increasing popularity of online social environments or platforms, such as within games \cite{paschke2021adolescent}, has led to a rise in toxic behavior \cite{kordyaka2023cycle, kordyaka2023exploring}, which can undermine these benefits \cite{cook2019whom}. A small percentage of players disproportionately contributes to the overall toxicity. For instance, in League of Legends, 1\% of the player base is responsible for approximately 5\% of the toxic speech \cite{stoop2019detecting}. Over time, such disruptive content can affect a large percentage of the player population \cite{frommel2022combating}, highlighting the critical need for effective moderation strategies.

\subsection{Moderation Approaches and Their Effectiveness}

Current approaches to combating toxicity in online gaming environments have largely focused on punitive measures. Studies of specific gaming communities, particularly League of Legends, have explored the use of permanent bans, automated moderation systems, and user reporting mechanisms \cite{kou2021punishment, kou2017code, kou2021flag}. However, these studies often suffer from low ecological validity, limiting the generalizability of their findings to broader gaming contexts.

In the broader context of online platforms, primarily social media, various studies have examined user experiences and policies around content moderation. For instance, a survey of over 900 users of commercially-moderated platforms revealed a paradox where user-moderated platforms, despite their greater transparency, are not perceived as less toxic compared to their commercially-moderated counterparts \cite{cook2021commercial}. Another study engaging 902 users across six social media platforms solicited opinions on effective countermeasures against toxicity, proposing strategies that remain largely untested for their efficacy in gaming contexts \cite{patel2021user}.

Specific user groups and platform contexts have also been explored in social media research, such as the moderation experiences of blind users on TikTok \cite{lyu2024got} and insights from volunteer moderators on Twitch \cite{cai2021moderation}. While these studies provide insights into moderation in online spaces, their direct applicability to gaming environments remains limited.

In the gaming context, a notable study explored harassment experiences among women in online video games, emphasizing the significant impact of both general and sexual harassment on women's participation and highlighting the pivotal role of the video game industry in addressing these challenges \cite{fox2017women}.

\subsection{Factors Influencing Moderation Effectiveness}

\subsubsection{Transparency and Fairness}

Recent gaming studies have identified strategies to maximize the efficacy of moderation. Transparency regarding when and why players are being moderated for toxic behavior, coupled with reinforcing good behavior, has been found to be effective \cite{lapolla2020tackling}. Additionally, research into player experiences in online multiplayer games highlights a significant demand for transparency and fairness in moderation processes \cite{ma2023transparency}.

\subsubsection{Timing of Interventions}

The timing of moderation actions has emerged as a critical factor in their effectiveness. While not specific to gaming, deterrence theory from criminology suggests that the certainty, severity, and celerity (swiftness) of punishment are key determinants in preventing undesirable behavior \cite{pratt2017empirical}. This aligns with findings from educational contexts, where postponing disciplinary actions can diminish their impact \cite{abramowitz1990effectiveness}.

\begin{figure}[t]
    \centering
    \includegraphics[width=1.0\textwidth]{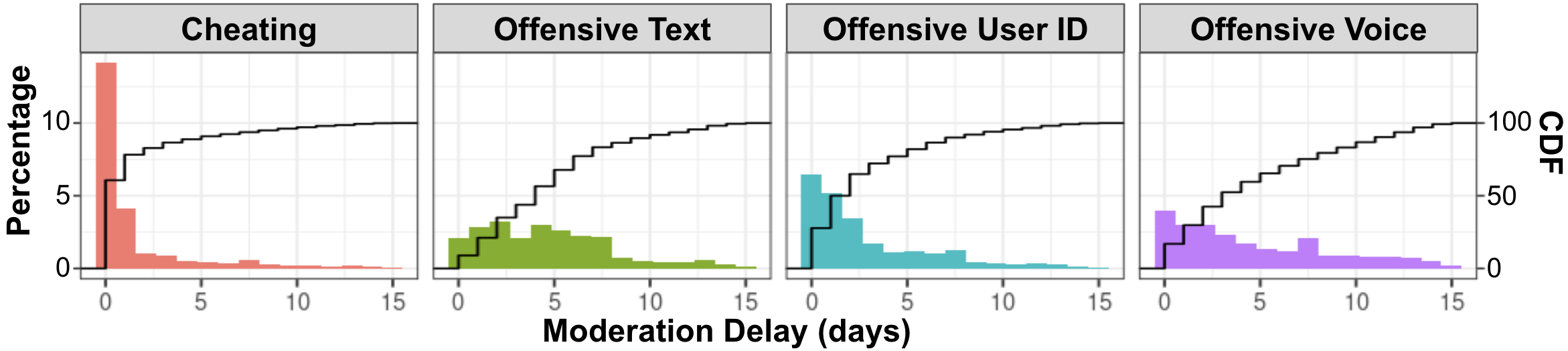}
    \caption{Distribution of Delay in Moderation; Figure showing both PDF and CDF of its distribution. Data generated from \textbf{Feb 1st, 2023} to \textbf{April 12th, 2023}.}
    \Description{Distribution of Delay in Moderation; Figure showing both PDF and CDF of its distribution.}
    \label{fig:mode_delay_hist}
\end{figure}

In social media moderation, delayed measures in content moderation tend to be less effective \cite{srinivasan2019content}. Studies have shown that action-effect delay can diminish an individual's sense of agency \cite{Wen2019-WENDDI}, which may apply to gaming contexts where players might disassociate their disruptive behavior from delayed punishment. In the context of \titleShort, moderation actions often come with a delay due to operational constraints. Figure \ref{fig:mode_delay_hist} illustrates the distribution of delays in moderation actions in our dataset.

These delays may occur due to the need to reach a specific threshold of reports before taking action, or reliance on human moderators for case review. Understanding the effects of these delays is crucial for optimizing moderation processes in gaming environments.

\subsubsection{Severity of Moderation Actions}

The severity of applied punishment is another crucial aspect of moderation. In gaming contexts, the expectation is that more severe punishment for disruptive behavior should result in more effective deterrence. Existing moderation approaches in online gaming often implement escalating punishment severity \cite{GameChan38:online}.

While not specific to gaming, behavioral economics research on ``loss aversion'' suggests that the immediate and certain prospect of punishment can have a more significant deterrent effect than the severity of the punishment itself \cite{kahneman2013prospect}. The application of this principle to gaming moderation warrants further investigation.

\subsection{Methodological Challenges in Studying Moderation Effects}

Studying the effects of moderation in real-world systems actively used by millions of users presents significant challenges \cite{kohavi2020trustworthy}. Random assignment of moderation is often ethically and practically infeasible. Recent studies, primarily in social media contexts, have attempted to address this challenge through various quasi-experimental designs and causal inference techniques \cite{ribeiro2023automated, wojck2022birdwatch, jimenezduran2022did, zhang2023china, jimenezduran2022economics}.

However, the application of these methodologies to online gaming contexts remains limited. A significant portion of the literature at the nexus of computer science (CS) and human-computer interaction (HCI) is based on correlational evidence from observational data or limited lab studies, which makes it difficult to isolate the assignment of treatment from its effects \cite{jimenezduran2022economics}.

\subsection{Gap in Literature and Study Motivation}

Despite the growing body of work on content moderation in online platforms, there remains a significant gap in our understanding of the causal effects of moderation on subsequent user behavior in online gaming environments. The complex interplay between moderation actions, their timing, severity, and their impact on both disruptive behavior and player engagement has not been comprehensively examined in a real-world gaming context.

Our study aims to address these gaps by employing a quasi-experimental design and causal inference techniques to examine the trade-offs between reducing disruptive behavior and player engagement when applying moderation in \titleShort. By focusing on the effects of moderation timing and severity on a set of four disruptive behaviors (cheating, offensive user name, chat, and voice), we seek to provide empirically grounded insights that can inform more effective and balanced moderation practices in online gaming environments.

This research not only contributes to the academic understanding of moderation effects but also has practical implications for game developers and community managers seeking to create safer and more enjoyable online gaming experiences.

\section{Hypotheses}

Based on the literature review, we formulate two primary research questions, each with associated hypotheses. These hypotheses are grounded in existing research on moderation in online environments, with a specific focus on their application to online gaming platforms like \titleShort.

\noindent \textbf{RQ1: What is the impact of moderation on players' disruptive behavior?}

\textbf{Hypothesis 1.1:} Moderation reduces disruptive behaviors among moderated players.

\textbf{Hypothesis 1.2:} \underline{Quick} moderation is more effective at reducing disruptive behaviors than delayed moderation\footnote{We define `Quick' moderation as occurring within three days of the offense, and compare it to moderation `Delayed' to a week after the offense. Refer to Section\ref{sec:design} for more detail.}.

\textbf{Hypothesis 1.3:} \underline{Stricter} moderation leads to a greater reduction in disruptive behaviors\footnote{Severity of moderation is determined by individual third-party moderators, who choose from a set of pre-defined actions. Refer to Figures \ref{table:moderation_actions} and \ref{table:moderation-actions-description} for more detail.}.

\textbf{Hypothesis 1.4:} \underline{Quick} implementation of \underline{stricter} moderation is more effective at reducing disruptive behaviors  compared to delayed implementation

\paragraph{\textbf{Motivation for hypotheses:}} These hypotheses are motivated by several strands of research in our literature review. \emph{Hypothesis 1.1} is grounded in the general expectation that moderation can effectively reduce toxic behavior, as suggested by studies in the gaming context \cite{lapolla2020tackling}. While most existing research focuses on social media platforms, we extend this expectation to online gaming environments. 
\emph{Hypotheses 1.2 and 1.4} are informed by deterrence theory \cite{pratt2017empirical} and findings from educational contexts \cite{abramowitz1990effectiveness}, which suggest that the swiftness of punishment is a key factor in its effectiveness. Although these studies are not specific to gaming, we apply this principle to our gaming context, hypothesizing that immediate moderation will be more effective than delayed moderation. \emph{Hypothesis 1.3} is based on the assumption underlying many existing moderation approaches in online gaming, which implement escalating punishment severity \cite{GameChan38:online}. This hypothesis also aligns with the general deterrence principle that more severe punishments should result in more effective deterrence \cite{klepper1989deterrent}. The interaction effect proposed in \emph{Hypothesis 1.4} combines the principles of immediacy and severity, suggesting that these factors may have a synergistic effect on reducing disruptive behavior.

\noindent \textbf{RQ2: What is the impact of moderation on participation rate?}

\textbf{Hypothesis 2.1:} Moderation reduces the participation rate among moderated players.

\textbf{Hypothesis 2.2:} \underline{Quick} moderation has a lesser negative impact on the participation rate compared to moderation applied with a delay.

\textbf{Hypothesis 2.3:} \underline{Stricter} moderation leads to a greater reduction in the participation rate than milder moderation.

\paragraph{\textbf{Motivation for hypotheses:}} These hypotheses address the potential dual impact of moderation on reducing disruptive behavior and on discouraging the participation of disruptive players in the first place, a critical consideration in online gaming environments. \emph{Hypothesis 2.1} is motivated by findings from studies like \cite{fox2017women}, which highlight potential negative consequences of moderation, such as decreased player participation. While moderation aims to create a better gaming environment, it may also discourage participation among moderated players. \emph{Hypothesis 2.2} extends the principle of immediacy to participation rates. We hypothesize that delayed moderation might lead to a more significant decrease in participation compared to immediate moderation. This is based on the idea that players might perceive delayed moderation as unfair or disconnected from their actions, potentially leading to frustration and disengagement. \emph{Hypothesis 2.3} is grounded in the concept of loss aversion from behavioral economics \cite{kahneman2013prospect}. While not directly studied in gaming contexts, we apply this principle to hypothesize that more severe moderation actions will lead to a greater reduction in participation rates. This hypothesis also considers the possibility that some players might engage in disruptive behavior purposefully \cite{lee2019disruptive} and might choose to stop playing rather than modify their behavior when faced with stricter moderation. These hypotheses collectively aim to explore the complex interplay between moderation actions, their timing and strictness, and their impact on both disruptive behavior and player participation in \titleShort. By testing these hypotheses, we seek to provide empirically grounded insights that can inform more effective and balanced moderation practices in online gaming environments, addressing the gap in current literature regarding the causal effects of moderation on player behavior in real-world gaming contexts.

\begin{figure}[tp!]
  \begin{center}
\includegraphics[width=0.8\textwidth]{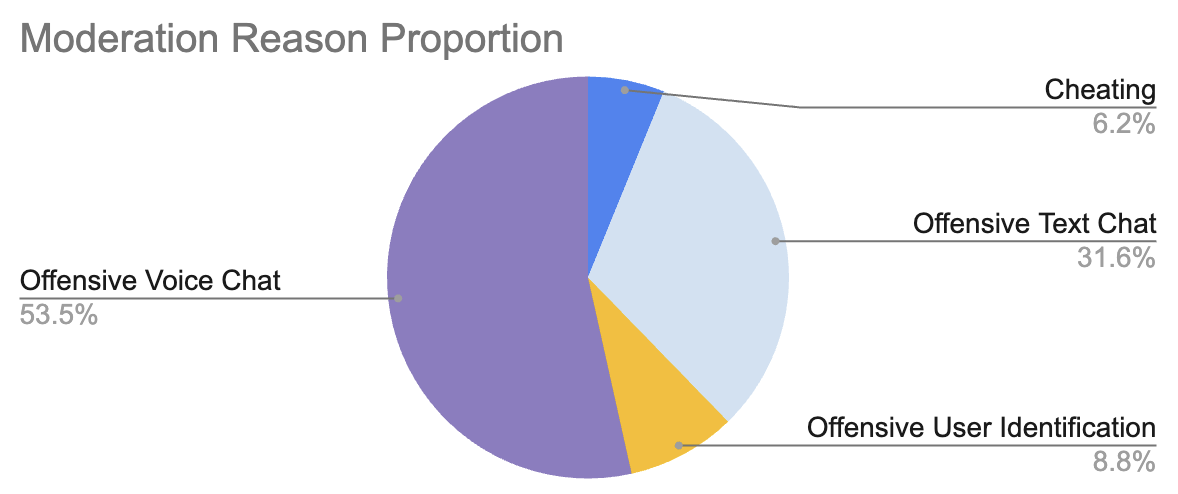}
  \caption{Breakdown of the unique proportion of players by moderation reason in our dataset. Data generated from \textbf{Feb 1st, 2023} to \textbf{April 12th, 2023}.}
  \Description{Breakdown of unique player proportion by moderation reason in our dataset: 53.5\% Offensive Voice Chat, 31.6\% offensive Text Chat, 8.8\% Offensive User ID, and 6.2\% Cheating.}
  \label{fig:mod_type_counts}
  \end{center}
\end{figure}

\section{Data}

In \titleShort, if players suspect other players are cheating or feel a player is using language or a name that they find offensive, players have the opportunity to report other players with the option of selecting corresponding report reasons. This might happen during or outside of a match. Moderation review is made after the report is generated and appropriate moderation actions can be taken.

If the disruptive behavior is confirmed, certain actions could be implemented based on the given category of moderation reason.\footnote{A detailed explanation on moderation in Call of Duty can be found in https://www.callofduty.com/blog/2024/01/call-of-duty-ricochet-modern-warfare-iii-warzone-anti-cheat-progress-report} There are four possible moderation reasons: cheating, offensive voice chat, offensive text chat, and offensive user ID. Figure~\ref{fig:mod_type_counts} summarizes the proportion of all possible moderation reasons in our dataset.
Table~\ref{tab:reason_action}, in supplementary materials, summarizes possible actions that could be taken given each moderation reason. 
Among each category of offensive type, there are multiple actions taken. These actions can be further broken down into \emph{Milder} and \emph{Stricter}, according to the severity of the applied punishment. Table~\ref{table:moderation_actions} shows the correspondence of actions and moderation reason, as well as the assigned severity of each action. We note that for the \emph{Cheating} disruptive behavior type, we were able to account for only one type of moderation action applied in our dataset - \textit{``Remove From Leaderboard''}, though other moderation actions exist. Hence, stratification into milder and stricter moderation is not possible (see Table~\ref{tab:reason_action}).

\subsection{Data Collection}
To answer the research questions, we use the player report data generated from Feb 1st, 2023 to April 12th, 2023. \footnote{For voice chat reports, we use March 21st to April 12th, for the voice chat moderation system was not in place before that. } The player data obtained for this analysis contains no demographic or individually identifying information. The database contains reported players, reporting players, and the reason for reporting. We also join the moderation records with the report data. The moderation record contains reported players, moderation date, reporting players associated with the action, and the specific actions taken. Note that one moderation action could target multiple reports and be directly linked to multiple reporting players. We then join these two tables to get reports with moderation information. We want to focus on players who are verified to have had disruptive behavior to avoid known issues with misreporting \cite{kou2021flag}. Therefore, for each player who is reported on day $T$, we only keep the sample if the player is moderated within 14 days and when the moderation can be directly linked to reports. In other words, denote the day they are moderated as $M$, we only keep those with $M \in [T, T + 14]$. To address cases when one moderation event targets multiple reports, we keep the first record of the report associated with each moderation record as the report date. A detailed description of the data pre-processing can be found in the Appendix \ref{apx:data_preprocessing}.

\begin{table}[t!]
\centering

\begin{NiceTabular}{@{}lp{10.3cm}l@{}}
\CodeBefore 
\Body
\toprule
\textbf{Offense Type} & \textbf{List of Moderation Actions Taken} & \textbf{Severity} \\ \midrule
Cheater & Remove From Leaderboard{} & \colorbox{Gray}{\makebox[3em]{N/A}} \\[5pt]

Offensive Text Chat & Warning Notice, Feature Flag & \colorbox{lime}{\makebox[3em]{\modMILD}} \\

Offensive Text Chat & \textbf{Penalty Notice}, Feature Flag & \colorbox{pink}{\makebox[3em]{\modSEVERE}} \\[5pt]

Offensive User ID & Rename User, Limit Allowed Renames, Update Clantag, Penalty Notice{} & \colorbox{lime}{\makebox[3em]{\modMILD}} \\
Offensive User ID & Rename User, Limit Allowed Renames, Update Clantag, Penalty Notice, \textbf{Feature Flag} & \colorbox{pink}{\makebox[3em]{\modSEVERE}} \\[15pt]

Offensive Voice Chat & Feature Flag & \colorbox{lime}{\makebox[3em]{\modMILD}} \\
Offensive Voice Chat & Feature Flag, \textbf{FeatureFlag}, Penalty Notice{} & \colorbox{pink}{\makebox[3em]{\modSEVERE}} \\ \bottomrule

\end{NiceTabular}
\caption{Classification of Moderation Actions Based on Severity. Several actions can be taken in response to particular disruptive player behavior. The actions categorized as Stricter are highlighted in bold.
}
\label{table:moderation_actions}
\end{table}

\subsection{Third Party Moderation}
\subsubsection{Industry Labels}

Table \ref{table:moderation-actions-description} provides a description for the meaning of each individual moderation action applied. We note that we don't control these actions as we would in an experimental setting, as they are applied in a live industry-scale moderation system. We can only log and observe these actions and their effects observationally. Additionally, we can collect the characteristics of the players both moderated or not.

\begin{table}[b!]
\centering
\begin{tabular}{lp{11.0cm}}
\hline
\textbf{Moderation Action} & \textbf{Description} \\
\hline
Remove From Leaderboard & 
A complete exclusion of a player’s account appearing on public leaderboards (generally as part of a temporary or permanent ban)
\\

Warning Notice & Notification warning the player to change their behavior \\

Penalty Notice & Notification to let the player know that their actions incurred some sort of penalty \\

Rename User & The user was renamed due to the offensive name tag (e.g. to something generic like User123) \\

Limit Allowed Renames & The \# of allowed renames for that user was adjusted (e.g. to 0 so they can't change their username anymore) \\

Update Clantag & The user's clantag was removed due to an offensive clan tag being used \\

Feature Flag & The user had a feature flag set on them (e.g. this could be a public voice mute) \\

\hline
\end{tabular}
\caption{Explanation of Industry Applied Moderation Actions}
\label{table:moderation-actions-description}
\end{table}

\subsubsection{Timing of Moderation} We further have no experimental control over the timing and selection of moderation actions. These are also selected based on the live industry-scale moderation system. As such, in this work, we operate within the confines of a quasi-experimental setup described in Section \ref{sec:design}. We can, however, observe how moderation actions are being applied and try to compensate for the systematic ways in which they are chosen based on observed player characteristics.

\section{Methodology}

\begin{figure}[t!]
  \begin{center}
\includegraphics[width=1.\textwidth]{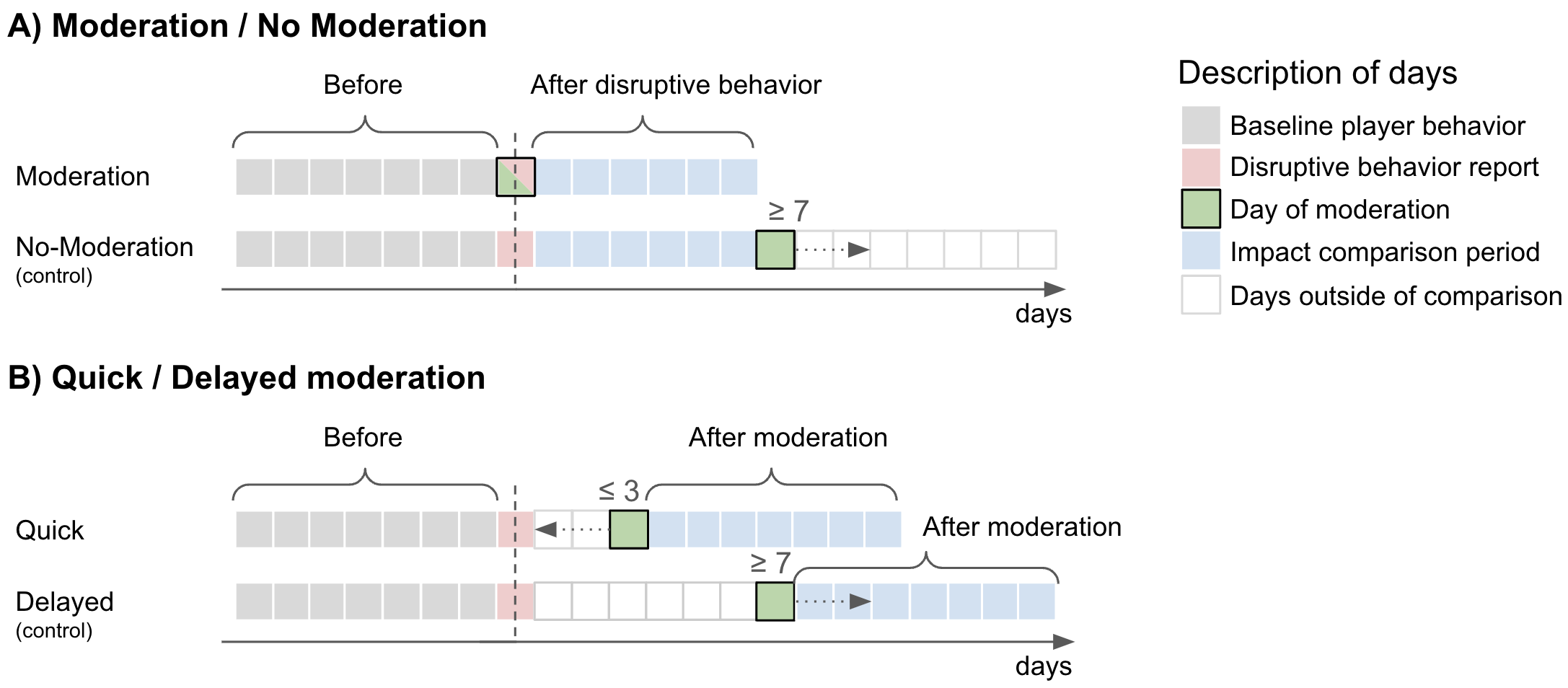}
  \caption{Overview of the quasi-experimental setup used for answering our research questions. \textbf{A)} Represents a setup where we compare the players moderated immediately to the ones moderated with substantial delay. Crucially the comparison period here is the player behavior post-disruptive event. This means that for the ``no-moderation'' group (control) it reflects how players behave when effectively being unmoderated. \textbf{B)} Represents a setup where we investigate the impact of swiftness of moderation action on post moderation player behaviors. In contrast to the previous setup, here both groups are compared after being exposed to moderation actions. Crucially, we include only players who were eventually moderated by the human moderation team to addresses the known issue of noise in player-generated reports, where players may be unfairly reported for cheating or toxic behavior \citep{beres2021don}.}
  \Description{}
  \label{fig:experiment_design_overview}
  \end{center}
\end{figure}

Our study employs a quasi-experimental design coupled with advanced causal machine learning (ML) techniques to estimate the impact of moderation versus no moderation on player behavior after engaging in disruptive behavior (Fig. \ref{fig:experiment_design_overview}-A). We further also investigate the impact of quick versus delayed moderation on player behavior post-moderation (Fig. \ref{fig:experiment_design_overview}-B). This section details our methodological approach, encompassing the design of treatment and control groups, outcome variable specification, causal inference framework, and model selection process.

\subsection{Quasi-Experimental Design}\label{sec:design}

The ideal examination of moderation effects and the impact of moderation delay would involve the random assignment of players to different experimental groups. However, such an experiment would necessitate allowing some players to engage in disruptive behavior without consequences, which is ethically and practically unfeasible in real-world gaming environments. Consequently, we employ a quasi-experimental approach, analyzing observational data where moderation actions have already been applied.
Our design leverages the natural variation in moderation timing within the game's moderation system. Let $T$ denote the date a player is reported and $M$ denote the date of moderation. We construct two distinct analyses:

\begin{enumerate}
\item \textbf{Moderation / No-moderation:} We compare players moderated on the same day as the report (treatment group, $M = T$) against those moderated with a substantial delay (control group, $M \geq T + 7$). Crucially, the comparison period here is the player behavior post-disruptive event. This means that for the ``no-moderation'' group (control) it reflects how players behave when effectively being unmoderated (Fig. \ref{fig:experiment_design_overview}-A).

\item \textbf{Quick / Delayed Moderation:} We compare players moderated within three days of the report (treatment group, $M \leq T + 3$) against those moderated with a substantial delay (control group, $M \geq T + 7$). Crucially, the comparison period here is the player behavior after moderation, effectively capturing the impact of moderation actions when applied quickly versus with delay (control). In contrast to the previous setup, here both groups are compared after being exposed to moderation actions (Fig. \ref{fig:experiment_design_overview}-B).
\end{enumerate}

This approach allows us to examine both 
important practical scenarios - the behavior of players during the period after being disruptive, but before being moderated, as well as the importance of moderation being swift for its effectiveness after application.

For the Moderation/No-moderation setup, we want to estimate the potential maximum impact of a highly efficient moderation system. As such, for the moderated group we focus on same-day moderation. On the other hand, to estimate player behavior without being moderated (control), we take players moderated with substantial delay. Crucially, we include only players who were eventually moderated by the human moderation team. This design choice addresses the known issue of noise in player-generated reports, where players may be unfairly reported for cheating or toxic behavior \citep{beres2021don}.

For the Quick/Delayed Moderation setup, the classification of moderation as quick is informed by the empirical distribution of moderation delays observed in our data, which exhibits a marked decline in moderation probability after the third day post-report across all offense types (as illustrated in Figure \ref{fig:mode_delay_hist}). This analysis also acknowledges the practical constraints faced by moderation teams. While same-day moderation is ideal, factors such as report volume, case complexity, and resource limitations may make it unfeasible in all instances. By extending the treatment window to three days, we capture a more achievable standard of quick response while still maintaining a clear distinction from delayed moderation.
 
By providing results for both these setups, our study can inform a range of policy decisions, from setting aspirational goals to defining achievable benchmarks for moderation systems. However, this approach introduces potential challenges, particularly in the analysis of moderation delay. Players moderated quickly may systematically differ from those moderated later; for instance, they might have committed more severe infractions. Conversely, delays might be caused by factors unrelated to the offense, such as moderator workload. These inherent disparities could undermine the reliability of using quasi-random timing of moderation actions for comparison.

To mitigate these concerns, we employ propensity score estimation \citep{rosenbaumRubin1983}, a technique widely used in observational studies for causal inference \citep{lechner2011}. This method allows us to balance the treatment and control groups based on observed confounders, creating a controlled virtual experiment to evaluate the effects of moderation timing. Specifically, we estimate the propensity score $e(X)$, defined as:
\begin{equation}
e(X) = P(W = 1 \mid X)
\end{equation}

where $W$ is the treatment indicator (1 for moderation or quick moderation, 0 for no moderation or delayed moderation) and $X$ is a vector of observed player characteristics and behavioral metrics preceding the report.
By conditioning on the propensity score, we can estimate the Average Treatment Effect (ATE) and Conditional Average Treatment Effect (CATE) of quick versus delayed moderation, effectively controlling for observable differences between the groups. This approach allows us to draw more robust causal inferences about the impact of moderation timing on player behavior and engagement.

\subsection{Outcome Variables}
We focus on two key outcome variables to quantify the impact of moderation timing:

\begin{enumerate}
\item Change in Report Rate ($\Delta R$):
    \begin{equation}
    \Delta R = \frac{\text{Reports Received}{w_1}}{\text{Matches Played}{w_1}} - \frac{\text{Reports Received}{w_0}}{\text{Matches Played}{w_0}}
    \end{equation}
    \item Change in Participation Rate ($\Delta P$):
    \begin{equation}
    \Delta P = \frac{\text{Days with Matches Played}_{w_1}}{7} - \frac{\text{Days with Matches Played}_{w_0}}{7}
    \end{equation}
\end{enumerate}

Here, $w_0 = [T-7, T-1]$ represents the week preceding the report, and $w_1 = [M, M+6]$ denotes the week following moderation. These metrics capture both the frequency of disruptive behavior and the level of player participation, providing a comprehensive view of moderation effects.

\subsection{Causal Inference Framework}

To estimate the causal effect of quick moderation, we employ a suite of causal machine learning techniques. Our primary estimates are the Average Treatment Effect (ATE) and the Conditional Average Treatment Effect (CATE):
\begin{align}
\text{ATE} &= \mathbb{E}[Y(1) - Y(0)] \\
\text{CATE}(X) &= \mathbb{E}[Y(1) - Y(0) \mid X]
\end{align}

where $Y(1)$ and $Y(0)$ represent the potential outcomes under quick and delayed moderation, respectively, and $X$ represents a vector of player characteristics.

The fundamental challenge in estimating these causal effects lies in the potential for confounding factors that influence both the timing of moderation and the outcomes of interest. To address this, we implement causal estimators, each with distinct approaches to controlling for confounding and estimating treatment effects.

\subsection{Causal Estimators}

We consider five primary causal estimators, summarized in Table \ref{tab:causal_estimators}. In this study, we primarily use the \emph{doubly robust} (DR) learner estimator for both the ATE and CATE. The DR Learner combines the propensity score model (PSM) and the outcome regression model (ORM) to offer robustness against potential model misspecifications \citep{sant2020doubly}. Crucially, the DR approach can yield unbiased estimates even if only one of these models is correctly specified, a feature particularly valuable in observational studies where strict model assumptions may not hold.

\begin{table}[b!]
\centering
\begin{tabular}{lp{6.0cm}p{6.5cm}}
\hline
\textbf{Estimator} & \textbf{Approach} & \textbf{Mathematical Formulation} \\
\hline
T-Learner & Separate models for treatment and control & $\hat{\tau}(X) = \hat{Y}_T(X) - \hat{Y}_C(X)$ \\
S-Learner & Single model with treatment as feature & $\hat{Y}(X, W) = f(X, W)$ \\
X-Learner & Cross-learning between treatment groups & Estimates treatment effects for both groups, then uses weighting scheme \\
R-Learner & Robinson transformation & Regresses outcome on covariates and treatment to get residuals, then regresses residuals on treatment \\
DR-Learner & Combines propensity score and regression & $\hat{\tau}(X) = \hat{\mu}_1(X) - \hat{\mu}_0(X) + \frac{W(Y - \hat{\mu}_1(X))}{\hat{e}(X)} - \frac{(1-W)(Y - \hat{\mu}_0(X))}{1-\hat{e}(X)}$ \\
\hline
\end{tabular}
\caption{Summary of Causal Machine Learning Estimators}
\label{tab:causal_estimators}
\end{table}

Consider players subjected to quick moderation (treatment group, $W=1$) or delayed moderation (control group, $W=0$). Let $X$ represent the covariates (player characteristics) and $Y$ be the outcome of interest (e.g., change in report rate or participation rate). The key components of our estimation framework are:
\begin{align}
\text{Propensity Score Model (PSM):} \quad e(X) &= P(W=1 \mid X) \\
\text{Outcome Regression Model (ORM):} \quad \mu(W, X) &= E[Y \mid W, X]
\end{align}

The DR-Learner estimators for ATE and CATE are given by:
\begin{align}
\widehat{\text{ATE}}_{\text{DR}} &= \frac{1}{N} \sum_{i=1}^N \left[ \frac{W_i Y_i - (W_i - e(X_i)) \hat{\mu}(1, X_i)}{e(X_i)} - \frac{(1-W_i) Y_i + (W_i - e(X_i)) \hat{\mu}(0, X_i)}{1 - e(X_i)} \right] \\
\widehat{\text{CATE}}_{\text{DR}}(X) &= \hat{\mu}(1, X) - \hat{\mu}(0, X) + \mathbb{E}\left[\frac{W(Y - \hat{\mu}(1, X))}{e(X)} - \frac{(1-W)(Y - \hat{\mu}(0, X))}{1-e(X)} \mid X\right]
\end{align}

The DR Learner is superior in handling heterogeneous treatment effects, allowing for a more nuanced understanding of how treatment impacts different subgroups. It can also incorporate non-linear relationships and interaction terms more effectively than traditional linear regression. Furthermore, the DR Learner offers flexibility in choosing different meta-learners for both components of the model, enabling the use of sophisticated machine learning algorithms when necessary.

\subsection{Model Selection and Robustness}

Our model selection process consists of two stages:

\begin{enumerate}
\item Meta-learner Selection: We compare the performance of all five estimators based on their ability to produce consistent and interpretable estimates of treatment effects.
\item Base Learner Selection: Within each meta-learner framework, we evaluate various machine learning algorithms as base learners, including Random Forests, XGBoost, and linear models.
\end{enumerate}

To assess the robustness of our results, we conduct a comprehensive comparison of CATE distributions across different meta-learners and base models (Figure \ref{fig:robust} in \ref{apx:model_selection}). This analysis reveals that while all estimators produce qualitatively similar results, the DR-Learner consistently yields the most stable and interpretable estimates across different specifications.

The DR-Learner emerges as our preferred specification for several reasons:

\begin{enumerate}
\item Robustness to Misspecification: The DR-Learner's double robustness property ensures consistent estimates if either the outcome model or the propensity score model is correctly specified.
\item Efficiency: It achieves the semiparametric efficiency bound under certain conditions, leading to more precise estimates.
\item Flexibility: The DR-Learner accommodates complex, non-linear relationships between covariates and outcomes, crucial for capturing the nuanced dynamics of player behavior.
\item Interpretability: It provides clear, interpretable estimates of both average and conditional treatment effects, facilitating meaningful insights into the heterogeneity of moderation impacts.
\end{enumerate}

Within the DR-Learner framework, we find that XGBoost performs optimally as the base learner for estimating the effect of moderation, while a linear model proves most effective for analyzing the effect of moderation delay (Figure \ref{fig:dr} in \ref{apx:model_selection}). This combination allows us to capture complex non-linear relationships where necessary while maintaining interpretability in simpler linear contexts.

Our models incorporate a comprehensive set of player characteristics as covariates:

\begin{equation}
X = \{\text{MatchScore}, \text{Assists}, \text{Eliminations}, \text{Deaths}, \text{DistanceTraveled}, \text{MoveSpeed}, \text{DamageDone}, \text{DamageTaken}, \text{Accuracy}\}
\end{equation}

These covariates serve dual purposes: they are used in the propensity score model to balance treatment and control groups, and as controls in the causal ML models to estimate heterogeneous treatment effects.

This framework allows us to estimate both the overall impact of moderation (ATE) and how this impact varies across different player subgroups (CATE), providing a nuanced understanding of moderation effectiveness. By leveraging the strengths of causal machine learning, particularly the DR-Learner, we can obtain robust estimates of treatment effects while accounting for the complex, potentially non-linear relationships between player characteristics, moderation timing, and outcomes of interest.

\subsection{Analysis of the Heterogeneity of Moderation Effects}

We investigate the heterogeneity of moderation effects across different player subpopulations, characterized by monitored performance metrics. Specifically, we analyze the Pearson correlations between individual Conditional Average Treatment Effects (CATE) and key player features, focusing on our key outcome metrics of repeated report and participation rates (Appendix~\ref{apx:heterogeneity}). To define subpopulations, we used several player performance metrics common in competitive action games \cite{karavolos2017learning, shim2011exploratory}, such as League of Legends (LoL) \cite{KilltoDe71:online, zhang2024sido}.

\begin{itemize}
    \item \emph{Average Match Score \textbf{(AMS)}} - provides a holistic view of player performance and is indicative of player skill and experience. This indicator is more aligned with team play and objectives and is often used for overall player ranking. It can include bonuses for objectives completed, which makes it a holistic measure of gameplay.

    \item \emph{Damage Skill Indicator \textbf{(DSI)}} - measures the ability to deal damage, which might not always correlate with kills. This metric is often used to evaluate a player’s offensive capabilities, particularly in games where assists and supportive play are valuable, such as LoL \cite{zhang2024sido}. It is defined as \( \frac{\text{average damage dealt}}{\text{average damage taken}} \).

    \item \emph{Kill-Death Ratio \textbf{(K/D)}} - focuses narrowly on combat effectiveness and measures individual success in combat. Commonly used as a benchmark for individual skill, particularly in competitive games, such as LoL \cite{KilltoDe71:online}. A higher K/D Ratio is generally seen as a sign of a skilled player. It is defined as \( \frac{\text{average kills}}{\text{average deaths}} \).

\end{itemize}

\section{Results}

\begin{figure}[!htp]
    \centering
    \includegraphics[width=\linewidth]{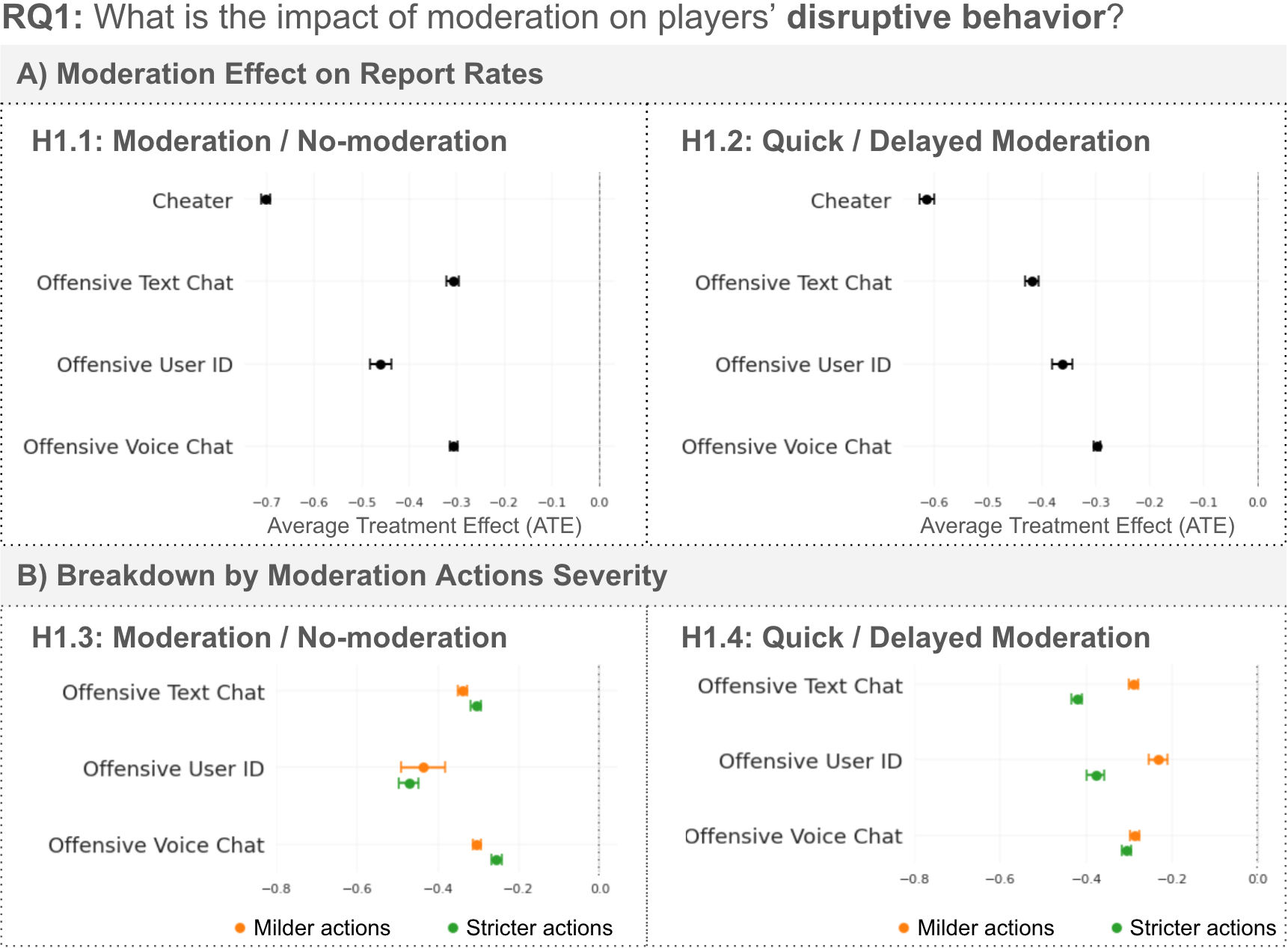}
    \caption{Analysis of the impact of moderation actions on \textbf{repeated report rates} (investigating RQ1), a measure of disruptive behavior. A) Effect of moderation/no-moderation (H1.1), quick moderation/delayed moderation (H1.2). B) Breakdown by action severity for moderation/no-moderation (H1.3) and quick moderation/delayed moderation (H1.4).}
    
    \Description{Analysis of the impact of moderation actions on repeated offense rates (RQ1), a measure of disruptive behavior. A) Effect of moderation/no-moderation (H1.1), quick moderation/delayed moderation (H1.2). B) Breakdown by action severity for moderation/no-moderation (H1.3) and quick moderation/delayed moderation (H1.4).}
    \label{fig:hypothesis1}
\end{figure}

\begin{figure}[t]
    \centering
    \includegraphics[width=\linewidth]{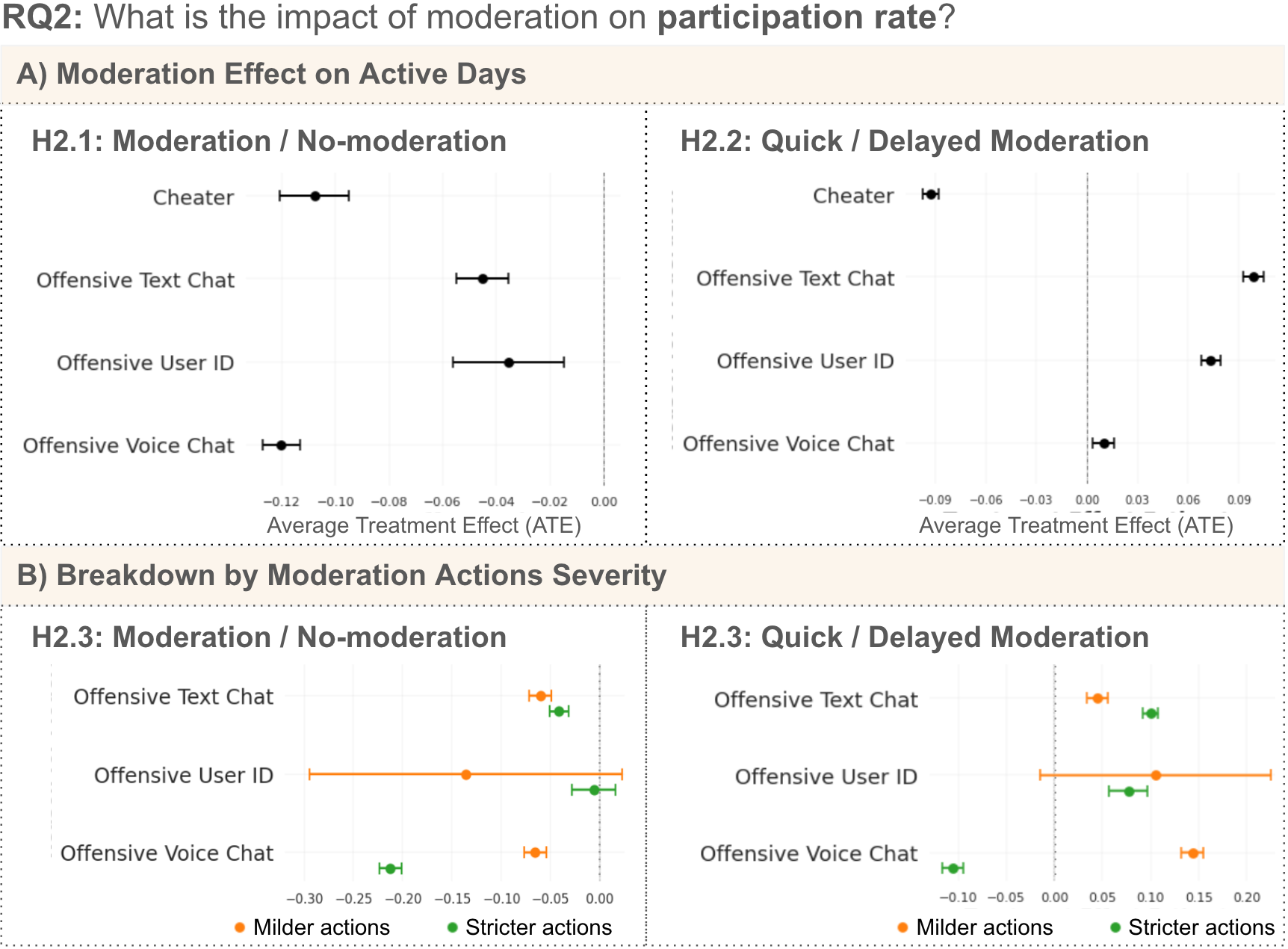}
    \caption{Testing hypotheses associated with RQ2,impact of moderation actions on \textbf{participation rates} (proportion of days with matches played). A) Effect of moderation/no-moderation (H2.1), quick moderation/delayed moderation (H2.2). A) Breakdown by action severity (H2.3).}
    \Description{Testing hypotheses associated with RQ2,  impact of moderation actions on \textbf{participation rates} (proportion of days with matches played). A) Effect of moderation/no-moderation (H2.1), quick moderation/delayed moderation (H2.2). A) Breakdown by action severity (H2.3).}
    \label{fig:hypothesis2}
\end{figure}

In Figure~\ref{fig:hypothesis1}, we present the estimated effects of moderation measures on altering player behavior, specifically regarding the changes in report rates about disruptive behavior (RQ1). Figure~\ref{fig:hypothesis2} reports the effect of moderation on player participation rate (RQ2).  In both graphs, we depict the effects of moderation vs no moderation, delayed moderation vs quick moderation, as well as the effects of different moderation action severity.   

\subsection{Impact of Moderation on Repeated Disruptive Behavior (RQ1)}

\paragraph{\textbf{Testing Hypothesis 1.1}}:
Moderation reduces disruptive behaviors among moderated players.

The evidence supports Hypothesis 1, demonstrating a clear decrease in disruptive behaviors among players who were moderated for various infractions. For instance, moderation targeted at \emph{Cheating} resulted in a substantial -70.33\% (95\% CI: -71.25\%, -69.41\%) reduction in subsequent report rates, indicating a significant deterrence effect. Similarly, moderation for \emph{Offensive Text Chat} and \emph{Offensive User ID} led to reductions in report rates by -30.84\% (95\% CI: -32.08\%, -20.60\%) and -46.05\% (95\% CI: -48.29\%, -43.81\%), respectively. These figures suggest that moderation effectively curtails repeated offenses (\textbf{Hypothesis 1.1 supported}).

\paragraph{\textbf{Testing Hypothesis 1.2:}}
Quick moderation is more effective at reducing disruptive behaviors than delayed moderation.

The evidence suggests a significant deterrent effect of quick moderation on repeat offenses (report rate): a -61.44\% (95\% CI: -62.78\%, -60.10\%) decrease in \emph{Cheating}, a -41.89\% (95\% CI: -43.05\%, -40.72\%) reduction in \emph{Offensive Text Chat} incidents, a -36.36\%  (95\% CI: -38.23\%, -34.49\%) decline in \emph{Offensive User ID} cases, and a -29.81\% (95\% CI: -30.51\%, -29.12\%) drop in \emph{Offensive Voice Chat} report rates. These findings indicate that quick moderation as compared to moderation with delay is more effective in curtailing disruptive behaviors on gaming platforms (\textbf{Hypothesis 1.2 supported}). 

\paragraph{\textbf{Testing Hypothesis 1.3:}}
Stricter moderation leads to a greater reduction in disruptive behaviors. 

Figure~\ref{fig:hypothesis1}-B further breaks down the effects of different actions among each moderation reason group according to Table~\ref{table:moderation_actions} into \emph{``Milder''} and \emph{``Stricter''} moderation actions. We can better understand the impact of these moderation strategies by looking at the variation in severity and the effect size. 
\emph{Stricter} moderation of \emph{Offensive Text Chat} results in a reduction of -30.58\%, 95\% CI (-33.02\%, -28.14\%)  while \emph{Milder} actions lead to a decrease of -33.91\%, with 95\% CI (-36.38\%, -31.44\%); with overlapping CIs suggesting no significant difference. Similarly for \emph{Offensive Voice Chat}, \emph{Stricter} actions resulted in a reduction of report rates by -25.48\% with 95\% CI (-28.14\%, -22.82\%) as compared to \emph{Milder} actions leading to higher reduction of-30.35\% with 95\% CI (-32.44\%, -28.26\%). 
Finally, for \emph{Offensive User ID} as well, the \emph{Stricter} moderation actions result in a reduction in report rates of -43.69\% with 95\% CI (-54.45\%, -32.92\%) comparable to \emph{Milder} actions at -47.11\% with 95\% CI (-51.93\%, -42.29\%). Evidence indicates that \emph{Stricter} and \emph{Milder} moderation actions have similar effects in reducing disruptive behavior as measured by a change in report rates (\textbf{Hypothesis 1.3 rejected}).  

\paragraph{\textbf{Testing Hypothesis 1.4:}} Quick implementation of stricter moderation is more effective at reducing disruptive behaviors compared to delayed implementation.

In Figure \ref{fig:hypothesis1} (H1.4) we can see that the differences between quick and delayed moderation actions are larger for \emph{stricter} compared to \emph{milder} actions. For \emph{Offensive Text Chat}, the report rate reduction is -42.21\% (95\% CI: -43.38\%, -41.04\%) for \emph{stricter} actions versus -29.04\% (95\% CI: -30.12\%, -27.96\%) for \emph{milder} actions; for \emph{Offensive User ID}, the reduction is at -37.83\% (95\% CI: -39.80\%, -35.85\%) for\emph{stricter} actions versus -23.28\% (95\% CI: -25.54\%, -21.03\%) for \emph{milder ones. Finally,} for \emph{Offensive Voice Chat}, the reduction in report rate is -30.68\% (95\% CI: -31.81\%, -29.54\%) for \emph{stricter} actions versus -28.72\% (95\% CI: -29.87\%, -27.57\%) for \emph{milder} ones. Across all these disruptive behavior categories, we can see the \emph{stricter} moderation actions taken \emph{quickly} are more effective in reducing repeated report rates. We can also see that all the action categories are more effective when taken quickly, regardless of the severity (\textbf{Hypothesis 1.4 supported}). 

\subsection{Impact of Moderation on Player Participation Rate (RQ2)}

The data from our study provides insights not only into the effectiveness of moderation in reducing disruptive behavior measured by report rates, but also into the impact it has on the participation rates of moderated players, measured by days with matches played. These results are depicted in Fig \ref{fig:hypothesis2}. We further test the hypotheses associated with impact of moderation on player participation rates.

\paragraph{\textbf{Testing Hypothesis 2.1:}}Moderation reduces the participation rate among moderated players.

Figure \ref{fig:hypothesis2} (H2.1) indeed shows that moderation has negative effect on participation rates of moderated players. Moderation for \emph{Cheating} was associated with a -10.80\% decline in participation, while moderation for \emph{Offensive Text Chat} and \emph{Offensive User ID} resulted in decreases of -4.53\% (95\% CI: -5.51\%, -3.55\%) and -3.57\% (95\% CI: -5.64\%, -1.50\%) in participation, respectively. Moreover, the largest drop in participation rates was observed following moderation for \emph{Offensive Voice Chat}, with a -12.03\% (95\% CI: -12.73\%, -11.32\%) reduction. These findings indicate that moderation also reduces the participation rates among those subjected to moderation actions (\textbf{Hypothesis 2.1 supported}).

\paragraph{\textbf{Testing Hypothesis 2.2:}} Quick moderation has a lesser negative impact on the participation rate compared to moderation applied with a delay.

In Figure \ref{fig:hypothesis2} (H2.2) we can see that while quick moderation for \emph{Cheating} led to a -9.28\% (95\% CI: -9.75\%, -8.81\%) decrease in participation rate of moderated players, quick actions against \emph{Offensive Text Chat} and \emph{Offensive User ID} were associated with increases in the relative  participation rates for the moderated players by 9.87\% (95\% CI: 9.25\%, 10.49\%) and 7.35\% (95\% CI: 6.79\%, 7.91\%), respectively. 
The minimal increase observed for \emph{Offensive Voice Chat} of 0.99\% (95\% CI: 0.37\%, 1.62\%) further suggests that the nature of the offense and player perceptions of moderation may influence participation rates. This represents a mixed impact, with quick moderation leading to both increases and decreases in participation rates depending on the offense type (\textbf{Hypothesis 2.2 partially supported}).

\paragraph{\textbf{Testing Hypothesis 2.3:}} Stricter moderation leads to a greater reduction in the participation rate than milder moderation.

Looking at Figure~\ref{fig:hypothesis2} (H2.3) we see that for \emph{Offensive Voice Chat}, stricter moderation actions results in a -21.2\% (95\% CI: -0.22.35\%, -20.19\%) decrease in player participation compared to milder actions at -6.57\% (95\% CI: -7.70\%, -5.45\%). Also, quick actions for \emph{Offensive Voice Chat} result in -10.60\% (95\% CI:-12.83\%, -8.38\%) higher reduction in player participation compared to delayed actions. However, these results don't hold for other offense types (\textbf{Hypothesis 2.3 partially supported}). \\

\subsection{Investigating the Heterogeneity of Moderation Effects}

\paragraph{\textbf{Heterogeneity of the Change in Report Rates:}}
Our findings indicate a strong negative correlation between CATE for \emph{report rates} and player performance indicators (Figure \ref{fig:cate_on_reports}). For \emph{\textbf{AMS}}, the correlations vary from $r=-0.30$ for \emph{Cheating}, to $r=-0.53$ for \emph{Offensive Text Chat} (Fig. \ref{fig:cate_on_reports}-a). For \emph{\textbf{DSI}}, we can observe a similar, but even more pronounced trend with $r=-0.36$ for \emph{Cheating} to $r=-0.57$ for \emph{Offensive Text Chat} (Fig. \ref{fig:cate_on_reports}-b). Finally, for \emph{\textbf{K/D}}, these negative correlations are the strongest and most consistent across disruptive behavior types, varying from $r=-0.50$ for \emph{Cheating} to $r=-0.62$ for \emph{Offensive Text Chat} (Fig. \ref{fig:cate_on_reports}-c). Higher-skilled players consistently exhibit greater reductions in disruptive behavior following moderation than less skilled players. Furthermore, these correlations are consistently the strongest for \emph{Offensive Text Chat}.

\paragraph{\textbf{Heterogeneity of the Change in Participation Rates:}}

We observe similar negative correlation between CATE for \emph{participations rates} and different player performance indicators (Figure \ref{fig:cate_on_participation}). For \emph{\textbf{AMS}}, the correlations vary from $r=-0.21$ for \emph{Cheating} to $r=-0.51$ for \emph{Offensive Text Chat} (Fig. \ref{fig:cate_on_participation}-a). An even stronger correlation can be observed for \emph{\textbf{DSI}} with $r=-0.48$ for \emph{Cheating} to $r=-0.63$ for \emph{Offensive Text Chat} (Fig. \ref{fig:cate_on_participation}-b); and for
\emph{\textbf{K/D}}, varying from $r=-0.58$ for \emph{Cheating} to $r=-0.67$ for \emph{Offensive Text Chat} (Fig. \ref{fig:cate_on_participation}-c). These trends have the same direction as for \emph{report rates}, but they are even more pronounced across all types of disruptive behaviors and player performance metrics.

\section{Discussion}

\subsection{Broader Impact}

\paragraph{\textbf{Real-world Insights and Ecological Validity:}} Our study offers a rare, large-scale examination of real-world moderation practices in a highly popular competitive action game, \titleShort. Unlike controlled experimental studies, our research benefits from the ecological validity of analyzing actual moderation actions applied to players in a live industry-scale gaming environment. This setting eliminates common experimental artifacts such as desirability bias \cite{nederhof1985methods}, where players alter their behavior to appear favorable in an experimental setting, placebo effects \cite{vaccaro2018illusion}, where behavior changes simply due to awareness of being observed, and selection bias \cite{cleave2010there}, where study participants do not represent the broader player population. By studying players who were not influenced by the knowledge of being observed, our findings maintain the integrity of player behavior and are directly applicable to real-world gaming contexts.

\paragraph{\textbf{Generalizability Across Game Genres:}} Our findings very likely extend beyond Call of Duty, providing valuable insights into the moderation of competitive action games broadly. This category is one of the most popular in gaming overall \cite{TopVideo36:online}, and it includes subgenres like first-person shooters (FPS), multiplayer online battle arenas (MOBA), and fighting games, which share similar competitive dynamics and forms of player communication \cite{migliore2021esports}. The types of retroactive, punitive moderation actions explored in our study are currently predominant in the gaming industry, aimed primarily at discouraging disruptive behaviors \cite{wijkstra2023help}. While alternative, prosocial and preventive forms of moderation are only now being proposed \cite{steinkuehler2023games}. Hence, the approaches examined in our work represent the most common real-world practices, making our results relevant across similar games \cite{wijkstra2023help}. However, the extent to which these findings generalize to less competitive genres like RPGs or sandbox games as well as to broader social media platforms remains an open question due to their differing gameplay and interaction dynamics. Yet, even for these settings, our results offer an important comparison benchmark for further work in the moderation space.

\subsection{Insights into Moderation Effectiveness}
Several observations from our analysis warrant further in-depth interpretation and considerations in future research on the moderation of disruptive player behavior.

\paragraph{\textbf{Different Moderation Impact on Cheaters vs. Toxic Players:}} Our study reveals distinct patterns in the effectiveness of moderation based on the nature of the offense. Actions against cheating resulted in a substantial reduction in subsequent repeated offense (measured by report rates), underscoring the effectiveness of deterrence in this context. In contrast, actions against toxic behaviors, such as offensive text and user IDs, also reduced report rates but with varying magnitudes. At the same time, the moderation of cheating, especially quick moderation of such behavior, led to a much more pronounced decrease in participation rates among these players compared to toxic players.
One explanation for such differences could be that the motivations underlying these disruptive behaviors differ significantly. Cheaters are driven by the desire to continue making progress in the game when they are stuck, or seek the thrill of control \cite{consalvo2009cheating}. Prior research also suggests that such behavior is more likely to occur if perceived risk of being caught is low \cite{chen2015group} or if such behavior is normalized within the game community \cite{doherty2014analysis}. On the other hand, the drive behind engaging in toxic behaviors is markedly different, with the main drivers such as pseudo-anonymity and lack of consequences \cite{lapidot2012effects}, retaliation cycles \cite{kordyaka2020towards}, subjectivity of toxic experience \cite{neto2017studying}, stress of competition \cite{kordyaka2020towards}, and personality traits \cite{kircaburun2018cyberbullying}. This indicates, that, at least a portion of the players engaging in toxic behaviors, but not cheating, may do so unintentionally by being unaware of community standards or are driven by temporary frustration. This divergence in motivations may explain why cheaters are more likely to disengage from the game following moderation compared to toxic players.

\paragraph{\textbf{Timing of Moderation:}} Our analysis emphasizes that quick moderation plays a pivotal role, especially in effectively reducing disruptive behavior across various offense types. For cheating, quick actions, as compared to delayed ones, significantly deter repeated offenses. As we proposed in our hypotheses, one explanation could be that the quick intervention enhances the perceived risk of consequences \cite{fragoso2014meet}
and disrupts the cycle of misconduct early \cite{grace2022policies}. 
We further found that a combination of both quick and stricter moderation is most effective in reducing repeated offenses. One explanation could be that such clear and quick moderation may help set proper community standards from the onset. Such community standards are part of the \emph{subjective norms} factor in \emph{Theory of Planned Behavior} \cite{ajzen2020theory} for behavior change, and hence can help explain the enhanced effectiveness of stricter and quick moderation. This suggests that prioritizing timely moderation can maintain a healthy gaming environment and strengthen community standards, particularly in contexts where players value rapid, transparent enforcement of rules.

\paragraph{\textbf{Dual Impact of Moderation:}} 
Our results suggest that moderation has a dual effect on disruptive players. First, it can reduce repeated offenses (as measured by report rates), and second, it can drive away players who seem unwilling to alter their disruptive behavior (as measured by participation rates). Both types of reactions to moderation are intuitive upon closer examination. Players who are willing to change their behavior after a nudge from moderation may not have realized they were being toxic, as the perception of toxicity can be subjective, even among seasoned players \cite{beres2021don}, and they might have lost the perceived pseudo-anonymity that fosters toxicity \cite{liu2023after}. On the other hand, players unwilling to change their behavior might have been primarily interested in a quick win motivated by the thrill of control \cite{consalvo2009cheating} or might have treated the game solely as an outlet for frustration \cite{saarinen2017toxic}. In these cases, realizing that such behavior would be met with strict consequences, they chose to stop playing, as their predominant motivation for engaging with the game was to be disruptive \cite{kou2020toxic}. This dual effect underscores moderation's role not just in behavior correction but also in platform health management. Yet, the findings highlight opportunities for designing interventions that persuade behavior change, particularly through mechanisms that encourage prosocial behaviors. However, for players whose primary intent is to disrupt, the deterrence effect of driving them away remains a valid outcome.

\paragraph{\textbf{Heterogeneity of Effects Based on Player Experience:}}

We found a correlation between player in-game performance and moderation effectiveness, with more skilled players being more reactive to moderation actions. This correlation is even more pronounced for performance indicators related to offensive capabilities and combat effectiveness rather than overall gameplay. One possible explanation could be that highly skilled players face greater visibility within the community and incur higher opportunity costs when penalized \cite{alvarado2022league}. Furthermore, for players with strong offensive capabilities, their aggressive playstyles and combat effectiveness often attract more attention, making their actions highly scrutinized \cite{egliston2013play}. This heightened visibility amplifies the social pressure to conform when moderated. For these players, the consequences of moderation can extend beyond gameplay, potentially affecting rankings or viewership if they are content creators \cite{felczak2023systemic}. These factors make skilled players more sensitive to moderation, significantly impacting both their behavior and participation rates. Another explanation could be that moderation feedback is perceived differently across skill levels \cite{beres2021don}. Higher-skilled players might better understand the specific reasons for their moderation \cite{kordyaka2023cycle}, leading to a more targeted and effective adjustment of their behavior. In contrast, less skilled players might not fully grasp the feedback's intent or relevance, resulting in less pronounced behavioral changes.

\paragraph{\textbf{Limitations of Observational Data and Use of Causal ML:}}

Our study utilizes real-world data from moderation practices within an industry-scale system, enhancing the ecological validity of our findings but also introducing certain limitations. Due to ethical constraints, we are unable to experimentally manipulate moderation actions \cite{kou2021flag}, necessitating the use of causal machine learning techniques to estimate the effectiveness of various moderation strategies \cite{kimmel2021causal}. While we employed state-of-the-art doubly robust machine learning methods \cite{scholkopf2022causality}, adjusted for a comprehensive set of observed player variables using propensity scoring \cite{shiba2021using}, and carefully constructed comparison groups \cite{west2000causal}, our approach still relies on the assumption that all relevant confounders are observed, which does not fully address the potential impact of unmeasured variables \cite{hatt2024sequential}. Additionally, we had no control over the types of moderation actions applied in response to different disruptive behaviors, which limits our ability to fully decouple the effects of specific moderation actions from the disruptive behaviors that triggered them. This introduces some complexity in distinguishing whether observed effects are attributable to the moderation action itself or to the nature of the disruptive behavior involved. Despite these limitations, our methods provide valuable practical insights that are difficult to achieve through other means \cite{de2021conceptualising}. Moreover, our findings can be tested and validated in future controlled experiments that aim to reproduce our observations, including those exploring less disruptive variations of moderation, such as changes in the timing or severity of interventions.

\subsection{Future Directions} Our work highlights several potential avenues for further research and practical interventions to better understand, mitigate, and address the complex and evolving nature of toxic behavior in online gaming environments. Future research could explore the use of adaptive moderation systems, such as bandit algorithms \cite{tucker2023bandits}, to dynamically select optimal interventions based on real-time feedback. Such systems could continuously refine moderation strategies to maximize their effectiveness and defer to human assistance when needed \cite{lykouris2024learning}. Further investigation is needed into how the presentation of moderation actions influences player behavior. The impact of transparency in moderation \cite{ma2023transparency} and the potential for spillover effects across different types of disruptive behaviors \cite{dong2020review,} warrant closer examination to develop comprehensive moderation strategies. Finally, targeted controlled experiments can be used to confirm the effects observed in this study. Future work could involve smaller-scale experiments with carefully selected participant subsets. These experiments could test specific hypotheses derived from our causal ML findings in a controlled setting, exploring variations in moderation timing, severity, and communication strategies.

\section{Conclusion}
This study employed a unique dataset from \titleFullName ~ featuring real-world, industry-scale data on gameplay and moderation activities. We estimated the causal impact of moderation on disruptive player behaviors, a key issue in the digital era. Our analysis provides insights into player interactions and  the efficacy of moderation strategies, revealing that while some players change behavior, others are driven away from the platform. Timely moderation, particularly for cheating, proved more effective than delayed actions. We also uncovered complex relationships between moderation effect, timing, and action severity, offering empirical support for prompt and decisive moderation. By leveraging this dataset, we make a significant contribution to understanding moderation's effectiveness in online gaming, offering both academic insights and practical guidelines for improving moderation strategies. Our findings serve as a reference for future research and practical applications in online behavior management.

\section{Acknowledgements}
We thank Gary Quan, Jonathan Lane, and Michael Vance for helpful comments.
Due to the player confidentiality and data privacy policies of \company, the data and code used in this paper cannot be made available by the researchers. 

\bibliographystyle{ACM-Reference-Format}
\bibliography{references}


\begin{thebibliography}{89}


\ifx \showCODEN    \undefined \def \showCODEN     #1{\unskip}     \fi
\ifx \showDOI      \undefined \def \showDOI       #1{#1}\fi
\ifx \showISBNx    \undefined \def \showISBNx     #1{\unskip}     \fi
\ifx \showISBNxiii \undefined \def \showISBNxiii  #1{\unskip}     \fi
\ifx \showISSN     \undefined \def \showISSN      #1{\unskip}     \fi
\ifx \showLCCN     \undefined \def \showLCCN      #1{\unskip}     \fi
\ifx \shownote     \undefined \def \shownote      #1{#1}          \fi
\ifx \showarticletitle \undefined \def \showarticletitle #1{#1}   \fi
\ifx \showURL      \undefined \def \showURL       {\relax}        \fi
\providecommand\bibfield[2]{#2}
\providecommand\bibinfo[2]{#2}
\providecommand\natexlab[1]{#1}
\providecommand\showeprint[2][]{arXiv:#2}

\bibitem[Abramowitz and O'Leary(1990)]%
        {abramowitz1990effectiveness}
\bibfield{author}{\bibinfo{person}{Ann~J Abramowitz} {and}
  \bibinfo{person}{Susan~G O'Leary}.} \bibinfo{year}{1990}\natexlab{}.
\newblock \showarticletitle{Effectiveness of delayed punishment in an applied
  setting}.
\newblock \bibinfo{journal}{\emph{Behavior Therapy}} \bibinfo{volume}{21},
  \bibinfo{number}{2} (\bibinfo{year}{1990}), \bibinfo{pages}{231--239}.
\newblock


\bibitem[Aguerri et~al\mbox{.}(2023)]%
        {aguerri2023enemy}
\bibfield{author}{\bibinfo{person}{Jes{\'u}s~C Aguerri}, \bibinfo{person}{Mario
  Santisteban}, {and} \bibinfo{person}{Fernando Mir{\'o}-Llinares}.}
  \bibinfo{year}{2023}\natexlab{}.
\newblock \showarticletitle{The Enemy Hates Best? Toxicity in League of Legends
  and Its Content Moderation Implications}.
\newblock \bibinfo{journal}{\emph{European Journal on Criminal Policy and
  Research}} \bibinfo{volume}{29}, \bibinfo{number}{3} (\bibinfo{year}{2023}),
  \bibinfo{pages}{437--456}.
\newblock


\bibitem[Ajzen(2020)]%
        {ajzen2020theory}
\bibfield{author}{\bibinfo{person}{Icek Ajzen}.}
  \bibinfo{year}{2020}\natexlab{}.
\newblock \showarticletitle{The theory of planned behavior: Frequently asked
  questions}.
\newblock \bibinfo{journal}{\emph{Human Behavior and Emerging Technologies}}
  \bibinfo{volume}{2}, \bibinfo{number}{4} (\bibinfo{year}{2020}),
  \bibinfo{pages}{314--324}.
\newblock


\bibitem[Alvarado and Arbaiza(2022)]%
        {alvarado2022league}
\bibfield{author}{\bibinfo{person}{Claudia Alvarado} {and}
  \bibinfo{person}{Francisco Arbaiza}.} \bibinfo{year}{2022}\natexlab{}.
\newblock \showarticletitle{League of Legends community's perception of
  influencer marketing from streamers on Twitch}. In
  \bibinfo{booktitle}{\emph{2022 17th Iberian Conference on Information Systems
  and Technologies (CISTI)}}. IEEE, \bibinfo{pages}{1--5}.
\newblock


\bibitem[Beres et~al\mbox{.}(2021)]%
        {beres2021don}
\bibfield{author}{\bibinfo{person}{Nicole~A Beres}, \bibinfo{person}{Julian
  Frommel}, \bibinfo{person}{Elizabeth Reid}, \bibinfo{person}{Regan~L
  Mandryk}, {and} \bibinfo{person}{Madison Klarkowski}.}
  \bibinfo{year}{2021}\natexlab{}.
\newblock \showarticletitle{Don’t You Know That You’re Toxic: Normalization
  of Toxicity in Online Gaming}. In \bibinfo{booktitle}{\emph{Proceedings of
  the 2021 CHI Conference on Human Factors in Computing Systems}} (, Yokohama,
  Japan,) \emph{(\bibinfo{series}{CHI '21})}. \bibinfo{publisher}{Association
  for Computing Machinery}, \bibinfo{address}{New York, NY, USA}, Article
  \bibinfo{articleno}{438}, \bibinfo{numpages}{15}~pages.
\newblock
\showISBNx{9781450380966}
\urldef\tempurl%
\url{https://doi.org/10.1145/3411764.3445157}
\showDOI{\tempurl}


\bibitem[Bourgonjon et~al\mbox{.}(2016)]%
        {bourgonjon2016players}
\bibfield{author}{\bibinfo{person}{Jeroen Bourgonjon}, \bibinfo{person}{Geert
  Vandermeersche}, \bibinfo{person}{Bram De~Wever}, \bibinfo{person}{Ronald
  Soetaert}, {and} \bibinfo{person}{Martin Valcke}.}
  \bibinfo{year}{2016}\natexlab{}.
\newblock \showarticletitle{Players’ perspectives on the positive impact of
  video games: A qualitative content analysis of online forum discussions}.
\newblock \bibinfo{journal}{\emph{New Media \& Society}} \bibinfo{volume}{18},
  \bibinfo{number}{8} (\bibinfo{year}{2016}), \bibinfo{pages}{1732--1749}.
\newblock


\bibitem[Cai et~al\mbox{.}(2021)]%
        {cai2021moderation}
\bibfield{author}{\bibinfo{person}{Jie Cai}, \bibinfo{person}{Donghee~Yvette
  Wohn}, {and} \bibinfo{person}{Mashael Almoqbel}.}
  \bibinfo{year}{2021}\natexlab{}.
\newblock \showarticletitle{Moderation Visibility: Mapping the Strategies of
  Volunteer Moderators in Live Streaming Micro Communities}. In
  \bibinfo{booktitle}{\emph{Proceedings of the 2021 ACM International
  Conference on Interactive Media Experiences}} (Virtual Event, USA)
  \emph{(\bibinfo{series}{IMX '21})}. \bibinfo{publisher}{Association for
  Computing Machinery}, \bibinfo{address}{New York, NY, USA},
  \bibinfo{pages}{61–72}.
\newblock
\showISBNx{9781450383899}
\urldef\tempurl%
\url{https://doi.org/10.1145/3452918.3458796}
\showDOI{\tempurl}


\bibitem[Canossa et~al\mbox{.}(2021)]%
        {canossa2021honor}
\bibfield{author}{\bibinfo{person}{Alessandro Canossa}, \bibinfo{person}{Dmitry
  Salimov}, \bibinfo{person}{Ahmad Azadvar}, \bibinfo{person}{Casper
  Harteveld}, {and} \bibinfo{person}{Georgios Yannakakis}.}
  \bibinfo{year}{2021}\natexlab{}.
\newblock \showarticletitle{For honor, for toxicity: Detecting toxic behavior
  through gameplay}.
\newblock \bibinfo{journal}{\emph{Proceedings of the ACM on Human-Computer
  Interaction}} \bibinfo{volume}{5}, \bibinfo{number}{CHI PLAY}
  (\bibinfo{year}{2021}), \bibinfo{pages}{1--29}.
\newblock


\bibitem[Chen and Wu(2015)]%
        {chen2015group}
\bibfield{author}{\bibinfo{person}{Vivian Hsueh~Hua Chen} {and}
  \bibinfo{person}{Yuehua Wu}.} \bibinfo{year}{2015}\natexlab{}.
\newblock \showarticletitle{Group identification as a mediator of the effect of
  players’ anonymity on cheating in online games}.
\newblock \bibinfo{journal}{\emph{Behaviour \& Information Technology}}
  \bibinfo{volume}{34}, \bibinfo{number}{7} (\bibinfo{year}{2015}),
  \bibinfo{pages}{658--667}.
\newblock


\bibitem[Cleave et~al\mbox{.}(2010)]%
        {cleave2010there}
\bibfield{author}{\bibinfo{person}{Blair~Llewellyn Cleave},
  \bibinfo{person}{Nikos Nikiforakis}, {and} \bibinfo{person}{Robert Slonim}.}
  \bibinfo{year}{2010}\natexlab{}.
\newblock \showarticletitle{Is there selection bias in laboratory experiments?}
\newblock \bibinfo{journal}{\emph{Univ. of Melbourne Dept. of Economics Working
  Paper}} \bibinfo{number}{1106} (\bibinfo{year}{2010}).
\newblock


\bibitem[Consalvo(2009)]%
        {consalvo2009cheating}
\bibfield{author}{\bibinfo{person}{Mia Consalvo}.}
  \bibinfo{year}{2009}\natexlab{}.
\newblock \bibinfo{booktitle}{\emph{Cheating: Gaining advantage in
  videogames}}.
\newblock \bibinfo{publisher}{mit press}.
\newblock


\bibitem[Cook et~al\mbox{.}(2019)]%
        {cook2019whom}
\bibfield{author}{\bibinfo{person}{Christine Cook}, \bibinfo{person}{Rianne
  Conijn}, \bibinfo{person}{Juli{\"e}tte Schaafsma}, {and}
  \bibinfo{person}{Marjolijn Antheunis}.} \bibinfo{year}{2019}\natexlab{}.
\newblock \showarticletitle{For whom the gamer trolls: A study of trolling
  interactions in the online gaming context}.
\newblock \bibinfo{journal}{\emph{Journal of Computer-Mediated Communication}}
  \bibinfo{volume}{24}, \bibinfo{number}{6} (\bibinfo{year}{2019}),
  \bibinfo{pages}{293--318}.
\newblock


\bibitem[Cook et~al\mbox{.}(2021)]%
        {cook2021commercial}
\bibfield{author}{\bibinfo{person}{Christine~L Cook}, \bibinfo{person}{Aashka
  Patel}, {and} \bibinfo{person}{Donghee~Yvette Wohn}.}
  \bibinfo{year}{2021}\natexlab{}.
\newblock \showarticletitle{Commercial versus volunteer: Comparing user
  perceptions of toxicity and transparency in content moderation across social
  media platforms}.
\newblock \bibinfo{journal}{\emph{Frontiers in Human Dynamics}}
  \bibinfo{volume}{3} (\bibinfo{year}{2021}), \bibinfo{pages}{626409}.
\newblock


\bibitem[Cullen and Kairam(2022)]%
        {cullen2022practicing}
\bibfield{author}{\bibinfo{person}{Amanda~LL Cullen} {and}
  \bibinfo{person}{Sanjay~R Kairam}.} \bibinfo{year}{2022}\natexlab{}.
\newblock \showarticletitle{Practicing moderation: Community moderation as
  reflective practice}.
\newblock \bibinfo{journal}{\emph{Proceedings of the ACM on Human-computer
  Interaction}} \bibinfo{volume}{6}, \bibinfo{number}{CSCW1}
  (\bibinfo{year}{2022}), \bibinfo{pages}{1--32}.
\newblock


\bibitem[de~Vocht et~al\mbox{.}(2021)]%
        {de2021conceptualising}
\bibfield{author}{\bibinfo{person}{Frank de Vocht},
  \bibinfo{person}{Srinivasa~Vittal Katikireddi}, \bibinfo{person}{Cheryl
  McQuire}, \bibinfo{person}{Kate Tilling}, \bibinfo{person}{Matthew Hickman},
  {and} \bibinfo{person}{Peter Craig}.} \bibinfo{year}{2021}\natexlab{}.
\newblock \showarticletitle{Conceptualising natural and quasi experiments in
  public health}.
\newblock \bibinfo{journal}{\emph{BMC medical research methodology}}
  \bibinfo{volume}{21} (\bibinfo{year}{2021}), \bibinfo{pages}{1--8}.
\newblock


\bibitem[Doherty et~al\mbox{.}(2014)]%
        {doherty2014analysis}
\bibfield{author}{\bibinfo{person}{Shawn~M Doherty}, \bibinfo{person}{Devin
  Liskey}, \bibinfo{person}{Christopher~M Via}, \bibinfo{person}{Christina
  Frederick}, \bibinfo{person}{Jason~P Kring}, {and} \bibinfo{person}{Dahai
  Liu}.} \bibinfo{year}{2014}\natexlab{}.
\newblock \showarticletitle{An analysis of expressed cheating behaviors in
  video games}. In \bibinfo{booktitle}{\emph{Proceedings of the Human Factors
  and Ergonomics Society Annual Meeting}}, Vol.~\bibinfo{volume}{58}. SAGE
  Publications Sage CA: Los Angeles, CA, \bibinfo{pages}{2393--2396}.
\newblock


\bibitem[Dong and Kelcey(2020)]%
        {dong2020review}
\bibfield{author}{\bibinfo{person}{Nianbo Dong} {and}
  \bibinfo{person}{Benjamin~M Kelcey}.} \bibinfo{year}{2020}\natexlab{}.
\newblock \bibinfo{title}{A review of Causality in a social world: Moderation,
  mediation, and spill-over}.
\newblock
\newblock


\bibitem[Egliston(2013)]%
        {egliston2013play}
\bibfield{author}{\bibinfo{person}{Benjamin Egliston}.}
  \bibinfo{year}{2013}\natexlab{}.
\newblock \showarticletitle{Play to win: How competitive modes of play have
  influenced cultural practice in digital games}.
\newblock  (\bibinfo{year}{2013}).
\newblock


\bibitem[Felczak(2023)]%
        {felczak2023systemic}
\bibfield{author}{\bibinfo{person}{Mateusz Felczak}.}
  \bibinfo{year}{2023}\natexlab{}.
\newblock \showarticletitle{Systemic issues with narratives of identity:
  Toxicity and esports media professionals}.
\newblock \bibinfo{journal}{\emph{Convergence}} \bibinfo{volume}{29},
  \bibinfo{number}{2} (\bibinfo{year}{2023}), \bibinfo{pages}{400--416}.
\newblock


\bibitem[Fox and Tang(2017)]%
        {fox2017women}
\bibfield{author}{\bibinfo{person}{Jesse Fox} {and} \bibinfo{person}{Wai~Yen
  Tang}.} \bibinfo{year}{2017}\natexlab{}.
\newblock \showarticletitle{Women’s experiences with general and sexual
  harassment in online video games: Rumination, organizational responsiveness,
  withdrawal, and coping strategies}.
\newblock \bibinfo{journal}{\emph{New media \& society}} \bibinfo{volume}{19},
  \bibinfo{number}{8} (\bibinfo{year}{2017}), \bibinfo{pages}{1290--1307}.
\newblock


\bibitem[Fragoso(2014)]%
        {fragoso2014meet}
\bibfield{author}{\bibinfo{person}{Suely Fragoso}.}
  \bibinfo{year}{2014}\natexlab{}.
\newblock \showarticletitle{Meet the HUEHUEs: A Sociotechnical Approach to
  Disruptive Behaviour in Multiplayer Online Games}.
\newblock \bibinfo{journal}{\emph{International Journal of Sociotechnology and
  Knowledge Development (IJSKD)}} \bibinfo{volume}{6}, \bibinfo{number}{3}
  (\bibinfo{year}{2014}), \bibinfo{pages}{26--44}.
\newblock


\bibitem[Frommel and Mandryk(2022)]%
        {frommel2022effective}
\bibfield{author}{\bibinfo{person}{Julian Frommel} {and} \bibinfo{person}{Regan
  Mandryk}.} \bibinfo{year}{2022}\natexlab{}.
\newblock \bibinfo{title}{Effective Toxicity Prediction in Online Multiplayer
  Gaming: Four Obstacles to Making Approaches Usable}.
\newblock \bibinfo{howpublished}{Mensch und Computer 2022 - Workshopband}.
\newblock
\urldef\tempurl%
\url{https://doi.org/10.18420/muc2022-mci-ws12-315}
\showDOI{\tempurl}


\bibitem[Frommel et~al\mbox{.}(2022)]%
        {frommel2022combating}
\bibfield{author}{\bibinfo{person}{Julian Frommel}, \bibinfo{person}{Regan~L.
  Mandryk}, {and} \bibinfo{person}{Madison Klarkowski}.}
  \bibinfo{year}{2022}\natexlab{}.
\newblock \showarticletitle{Challenges to Combating Toxicity and Harassment in
  Multiplayer Games: Involving the HCI Games Research Community}. In
  \bibinfo{booktitle}{\emph{Extended Abstracts of the 2022 Annual Symposium on
  Computer-Human Interaction in Play}} (Bremen, Germany)
  \emph{(\bibinfo{series}{CHI PLAY '22})}. \bibinfo{publisher}{Association for
  Computing Machinery}, \bibinfo{address}{New York, NY, USA},
  \bibinfo{pages}{263–265}.
\newblock
\showISBNx{9781450392112}
\urldef\tempurl%
\url{https://doi.org/10.1145/3505270.3558359}
\showDOI{\tempurl}


\bibitem[GGWP(2023)]%
        {GameChan38:online}
\bibfield{author}{\bibinfo{person}{GGWP}.} \bibinfo{year}{2023}\natexlab{}.
\newblock \bibinfo{title}{Game Changer: Impact of Chat Sanctions on Toxicity -
  GGWP - the first AI-powered game moderation platform}.
\newblock
  \bibinfo{howpublished}{\url{https://www.ggwp.com/blog/game-changer-impact-of-chat-sanctions-on-toxicity/}}.
\newblock
\newblock
\shownote{(Accessed on 02/11/2024)}.


\bibitem[Ghosh(2021)]%
        {ghosh2021analyzing}
\bibfield{author}{\bibinfo{person}{Ayushi Ghosh}.}
  \bibinfo{year}{2021}\natexlab{}.
\newblock \showarticletitle{Analyzing Toxicity in Online Gaming Communities}.
\newblock \bibinfo{journal}{\emph{Turkish Journal of Computer and Mathematics
  Education (TURCOMAT)}} \bibinfo{volume}{12}, \bibinfo{number}{10}
  (\bibinfo{year}{2021}), \bibinfo{pages}{4448--4455}.
\newblock


\bibitem[Gorwa et~al\mbox{.}(2020)]%
        {gorwa2020algorithmic}
\bibfield{author}{\bibinfo{person}{Robert Gorwa}, \bibinfo{person}{Reuben
  Binns}, {and} \bibinfo{person}{Christian Katzenbach}.}
  \bibinfo{year}{2020}\natexlab{}.
\newblock \showarticletitle{Algorithmic content moderation: Technical and
  political challenges in the automation of platform governance}.
\newblock \bibinfo{journal}{\emph{Big Data \& Society}} \bibinfo{volume}{7},
  \bibinfo{number}{1} (\bibinfo{year}{2020}),
  \bibinfo{pages}{2053951719897945}.
\newblock


\bibitem[Grace et~al\mbox{.}(2022)]%
        {grace2022policies}
\bibfield{author}{\bibinfo{person}{Thomas~D Grace}, \bibinfo{person}{Ian
  Larson}, {and} \bibinfo{person}{Katie Salen}.}
  \bibinfo{year}{2022}\natexlab{}.
\newblock \showarticletitle{Policies of misconduct: a content analysis of codes
  of conduct for online multiplayer games}.
\newblock \bibinfo{journal}{\emph{Proceedings of the ACM on Human-Computer
  Interaction}} \bibinfo{volume}{6}, \bibinfo{number}{CHI PLAY}
  (\bibinfo{year}{2022}), \bibinfo{pages}{1--23}.
\newblock


\bibitem[Hatt and Feuerriegel(2024)]%
        {hatt2024sequential}
\bibfield{author}{\bibinfo{person}{Tobias Hatt} {and} \bibinfo{person}{Stefan
  Feuerriegel}.} \bibinfo{year}{2024}\natexlab{}.
\newblock \showarticletitle{Sequential deconfounding for causal inference with
  unobserved confounders}. In \bibinfo{booktitle}{\emph{Causal Learning and
  Reasoning}}. PMLR, \bibinfo{pages}{934--956}.
\newblock


\bibitem[Horta~Ribeiro et~al\mbox{.}(2023)]%
        {ribeiro2023automated}
\bibfield{author}{\bibinfo{person}{Manoel Horta~Ribeiro},
  \bibinfo{person}{Justin Cheng}, {and} \bibinfo{person}{Robert West}.}
  \bibinfo{year}{2023}\natexlab{}.
\newblock \showarticletitle{Automated Content Moderation Increases Adherence to
  Community Guidelines}.
\newblock \bibinfo{journal}{\emph{Proceedings of the ACM Web Conference 2023}}
  \bibinfo{volume}{1} (\bibinfo{year}{2023}), \bibinfo{pages}{2666–2676}.
\newblock
\showISBNx{9781450394161}
\urldef\tempurl%
\url{https://doi.org/10.1145/3543507.3583275}
\showDOI{\tempurl}


\bibitem[Jiménez-Durán(2022)]%
        {jimenezduran2022economics}
\bibfield{author}{\bibinfo{person}{Rafael Jiménez-Durán}.}
  \bibinfo{year}{2022}\natexlab{}.
\newblock \bibinfo{booktitle}{\emph{The economics of content moderation: Theory
  and experimental evidence from hate speech on Twitter}}.
\newblock \bibinfo{type}{{T}echnical {R}eport} 324.
  \bibinfo{institution}{University of Chicago Booth School of Business, Stigler
  Center for the Study of the Economy and the State},
  \bibinfo{address}{Chicago, IL}.
\newblock


\bibitem[Jiménez~Durán et~al\mbox{.}(2023)]%
        {jimenezduran2022did}
\bibfield{author}{\bibinfo{person}{Rafael Jiménez~Durán},
  \bibinfo{person}{Karsten Müller}, {and} \bibinfo{person}{Carlo Schwarz}.}
  \bibinfo{year}{2023}\natexlab{}.
\newblock \showarticletitle{The Effect of Content Moderation on Online and
  Offline Hate: Evidence from Germany's NetzDG}.
\newblock \bibinfo{journal}{\emph{Available at SSRN 4230296}}
  \bibinfo{volume}{1} (\bibinfo{year}{2023}), \bibinfo{pages}{1--75}.
\newblock
\urldef\tempurl%
\url{http://dx.doi.org/10.2139/ssrn.4230296}
\showURL{%
\tempurl}


\bibitem[Kaddour et~al\mbox{.}(2022)]%
        {kaddour2022causal}
\bibfield{author}{\bibinfo{person}{Jean Kaddour}, \bibinfo{person}{Aengus
  Lynch}, \bibinfo{person}{Qi Liu}, \bibinfo{person}{Matt~J Kusner}, {and}
  \bibinfo{person}{Ricardo Silva}.} \bibinfo{year}{2022}\natexlab{}.
\newblock \bibinfo{title}{Causal machine learning: A survey and open problems}.
\newblock , \bibinfo{numpages}{arXiv--2206}~pages.
\newblock


\bibitem[Kahneman and Tversky(2013)]%
        {kahneman2013prospect}
\bibfield{author}{\bibinfo{person}{Daniel Kahneman} {and} \bibinfo{person}{Amos
  Tversky}.} \bibinfo{year}{2013}\natexlab{}.
\newblock \showarticletitle{Prospect theory: An analysis of decision under
  risk}.
\newblock In \bibinfo{booktitle}{\emph{Handbook of the fundamentals of
  financial decision making: Part I}}. \bibinfo{publisher}{World Scientific},
  \bibinfo{address}{Singapore}, \bibinfo{pages}{99--127}.
\newblock


\bibitem[Karavolos et~al\mbox{.}(2017)]%
        {karavolos2017learning}
\bibfield{author}{\bibinfo{person}{Daniel Karavolos}, \bibinfo{person}{Antonios
  Liapis}, {and} \bibinfo{person}{Georgios Yannakakis}.}
  \bibinfo{year}{2017}\natexlab{}.
\newblock \showarticletitle{Learning the patterns of balance in a multi-player
  shooter game}. In \bibinfo{booktitle}{\emph{Proceedings of the 12th
  international conference on the foundations of digital games}}.
  \bibinfo{pages}{1--10}.
\newblock


\bibitem[Kimmel et~al\mbox{.}(2021)]%
        {kimmel2021causal}
\bibfield{author}{\bibinfo{person}{Kaitlin Kimmel}, \bibinfo{person}{Laura~E
  Dee}, \bibinfo{person}{Meghan~L Avolio}, {and} \bibinfo{person}{Paul~J
  Ferraro}.} \bibinfo{year}{2021}\natexlab{}.
\newblock \showarticletitle{Causal assumptions and causal inference in
  ecological experiments}.
\newblock \bibinfo{journal}{\emph{Trends in Ecology \& Evolution}}
  \bibinfo{volume}{36}, \bibinfo{number}{12} (\bibinfo{year}{2021}),
  \bibinfo{pages}{1141--1152}.
\newblock


\bibitem[K{\i}rcaburun and Tosunta{\c{s}}(2018)]%
        {kircaburun2018cyberbullying}
\bibfield{author}{\bibinfo{person}{Ka{\u{g}}an K{\i}rcaburun} {and}
  \bibinfo{person}{{\c{S}}ule~Bet{\"u}l Tosunta{\c{s}}}.}
  \bibinfo{year}{2018}\natexlab{}.
\newblock \showarticletitle{Cyberbullying perpetration among undergraduates:
  Evidence of the roles of chronotype and sleep quality}.
\newblock \bibinfo{journal}{\emph{Biological rhythm research}}
  \bibinfo{volume}{49}, \bibinfo{number}{2} (\bibinfo{year}{2018}),
  \bibinfo{pages}{247--265}.
\newblock


\bibitem[Klepper and Nagin(1989)]%
        {klepper1989deterrent}
\bibfield{author}{\bibinfo{person}{Steven Klepper} {and}
  \bibinfo{person}{Daniel Nagin}.} \bibinfo{year}{1989}\natexlab{}.
\newblock \showarticletitle{The deterrent effect of perceived certainty and
  severity of punishment revisited}.
\newblock \bibinfo{journal}{\emph{Criminology}} \bibinfo{volume}{27},
  \bibinfo{number}{4} (\bibinfo{year}{1989}), \bibinfo{pages}{721--746}.
\newblock


\bibitem[Kocielnik et~al\mbox{.}(2023)]%
        {kocielnik6challenges}
\bibfield{author}{\bibinfo{person}{Rafal Kocielnik}, \bibinfo{person}{Zhuofang
  Li}, \bibinfo{person}{Claudia Kann}, \bibinfo{person}{Deshawn Sambrano},
  \bibinfo{person}{Jacob Morrier}, \bibinfo{person}{Mitchell Linegar},
  \bibinfo{person}{Carly Taylor}, \bibinfo{person}{Min Kim},
  \bibinfo{person}{Nabiha Naqvie}, \bibinfo{person}{Feri Soltani},
  \bibinfo{person}{Arman Dehpanah}, \bibinfo{person}{Grant Cahill},
  \bibinfo{person}{Animashree Anandkumar}, {and} \bibinfo{person}{R.~Michael
  Alvarez}.} \bibinfo{year}{2023}\natexlab{}.
\newblock \showarticletitle{Challenges in Moderating Disruptive Player Behavior
  in Online Competitive Action Games}.
\newblock \bibinfo{journal}{\emph{Frontiers in Computer Science}}
  \bibinfo{volume}{6} (\bibinfo{year}{2023}), \bibinfo{pages}{1283735}.
\newblock


\bibitem[Kohavi et~al\mbox{.}(2020)]%
        {kohavi2020trustworthy}
\bibfield{author}{\bibinfo{person}{Ron Kohavi}, \bibinfo{person}{Diane Tang},
  {and} \bibinfo{person}{Ya Xu}.} \bibinfo{year}{2020}\natexlab{}.
\newblock \bibinfo{booktitle}{\emph{Trustworthy online controlled experiments:
  A practical guide to a/b testing}}.
\newblock \bibinfo{publisher}{Cambridge University Press},
  \bibinfo{address}{Cambridge, UK}.
\newblock


\bibitem[Kordyaka et~al\mbox{.}(2020)]%
        {kordyaka2020towards}
\bibfield{author}{\bibinfo{person}{Bastian Kordyaka},
  \bibinfo{person}{Katharina Jahn}, {and} \bibinfo{person}{Bjoern Niehaves}.}
  \bibinfo{year}{2020}\natexlab{}.
\newblock \showarticletitle{Towards a unified theory of toxic behavior in video
  games}.
\newblock \bibinfo{journal}{\emph{Internet Research}} \bibinfo{volume}{30},
  \bibinfo{number}{4} (\bibinfo{year}{2020}), \bibinfo{pages}{1081--1102}.
\newblock


\bibitem[Kordyaka and Kruse(2021)]%
        {kordyaka2021curing}
\bibfield{author}{\bibinfo{person}{Bastian Kordyaka} {and}
  \bibinfo{person}{Bj{\"o}rn Kruse}.} \bibinfo{year}{2021}\natexlab{}.
\newblock \showarticletitle{Curing toxicity--developing design principles to
  buffer toxic behaviour in massive multiplayer online games}.
\newblock \bibinfo{journal}{\emph{Safer communities}} \bibinfo{volume}{20},
  \bibinfo{number}{3} (\bibinfo{year}{2021}), \bibinfo{pages}{133--149}.
\newblock


\bibitem[Kordyaka et~al\mbox{.}(2023a)]%
        {kordyaka2023cycle}
\bibfield{author}{\bibinfo{person}{Bastian Kordyaka}, \bibinfo{person}{Samuli
  Laato}, \bibinfo{person}{Katharina Jahn}, \bibinfo{person}{Juho Hamari},
  {and} \bibinfo{person}{Bjoern Niehaves}.} \bibinfo{year}{2023}\natexlab{a}.
\newblock \showarticletitle{The Cycle of Toxicity: Exploring Relationships
  between Personality and Player Roles in Toxic Behavior in Multiplayer Online
  Battle Arena Games}.
\newblock \bibinfo{journal}{\emph{Proceedings of the ACM on Human-Computer
  Interaction}} \bibinfo{volume}{7}, \bibinfo{number}{CHI PLAY}
  (\bibinfo{year}{2023}), \bibinfo{pages}{611--641}.
\newblock


\bibitem[Kordyaka et~al\mbox{.}(2023b)]%
        {kordyaka2023exploring}
\bibfield{author}{\bibinfo{person}{Bastian Kordyaka}, \bibinfo{person}{Solip
  Park}, \bibinfo{person}{Jeanine Krath}, {and} \bibinfo{person}{Samuli
  Laato}.} \bibinfo{year}{2023}\natexlab{b}.
\newblock \showarticletitle{Exploring the relationship between offline cultural
  environments and toxic behavior tendencies in multiplayer online games}.
\newblock \bibinfo{journal}{\emph{ACM Transactions on Social Computing}}
  \bibinfo{volume}{6}, \bibinfo{number}{1-2} (\bibinfo{year}{2023}),
  \bibinfo{pages}{1--20}.
\newblock


\bibitem[Kou(2020)]%
        {kou2020toxic}
\bibfield{author}{\bibinfo{person}{Yubo Kou}.} \bibinfo{year}{2020}\natexlab{}.
\newblock \showarticletitle{Toxic behaviors in team-based competitive gaming:
  The case of league of legends}. In \bibinfo{booktitle}{\emph{Proceedings of
  the annual symposium on computer-human interaction in play}}.
  \bibinfo{pages}{81--92}.
\newblock


\bibitem[Kou(2021)]%
        {kou2021punishment}
\bibfield{author}{\bibinfo{person}{Yubo Kou}.} \bibinfo{year}{2021}\natexlab{}.
\newblock \showarticletitle{Punishment and Its Discontents: An Analysis of
  Permanent Ban in an Online Game Community}.
\newblock \bibinfo{journal}{\emph{Proceedings of the ACM on Human-Computer
  Interaction}} \bibinfo{volume}{5}, \bibinfo{number}{CSCW2}
  (\bibinfo{year}{2021}), \bibinfo{pages}{1--21}.
\newblock


\bibitem[Kou and Gui(2017)]%
        {kou2017code}
\bibfield{author}{\bibinfo{person}{Yubo Kou} {and} \bibinfo{person}{Xinning
  Gui}.} \bibinfo{year}{2017}\natexlab{}.
\newblock \showarticletitle{When code governs community}. In
  \bibinfo{booktitle}{\emph{50th Annual Hawaii International Conference on
  System Sciences, HICSS 2017}}, Vol.~\bibinfo{volume}{1}. IEEE Computer
  Society, \bibinfo{publisher}{IEEE}, \bibinfo{address}{Hawaii, USA},
  \bibinfo{pages}{2056--2064}.
\newblock
\urldef\tempurl%
\url{https://api.semanticscholar.org/CorpusID:37012220}
\showURL{%
\tempurl}


\bibitem[Kou and Gui(2021)]%
        {kou2021flag}
\bibfield{author}{\bibinfo{person}{Yubo Kou} {and} \bibinfo{person}{Xinning
  Gui}.} \bibinfo{year}{2021}\natexlab{}.
\newblock \showarticletitle{Flag and flaggability in automated moderation: The
  case of reporting toxic behavior in an online game community}.
\newblock \bibinfo{journal}{\emph{Proceedings of the 2021 CHI Conference on
  Human Factors in Computing Systems}}  \bibinfo{volume}{1}
  (\bibinfo{year}{2021}), \bibinfo{pages}{1--12}.
\newblock


\bibitem[Kriz(2020)]%
        {kriz2020gaming}
\bibfield{author}{\bibinfo{person}{Willy~C Kriz}.}
  \bibinfo{year}{2020}\natexlab{}.
\newblock \showarticletitle{Gaming in the Time of COVID-19}.
\newblock \bibinfo{journal}{\emph{Simulation \& Gaming}} \bibinfo{volume}{51},
  \bibinfo{number}{4} (\bibinfo{year}{2020}), \bibinfo{pages}{403--410}.
\newblock


\bibitem[Kwak et~al\mbox{.}(2015)]%
        {kwak2015exploring}
\bibfield{author}{\bibinfo{person}{Haewoon Kwak}, \bibinfo{person}{Jeremy
  Blackburn}, {and} \bibinfo{person}{Seungyeop Han}.}
  \bibinfo{year}{2015}\natexlab{}.
\newblock \showarticletitle{Exploring Cyberbullying and Other Toxic Behavior in
  Team Competition Online Games}. In \bibinfo{booktitle}{\emph{Proceedings of
  the 33rd Annual ACM Conference on Human Factors in Computing Systems}}
  (Seoul, Republic of Korea) \emph{(\bibinfo{series}{CHI '15})}.
  \bibinfo{publisher}{Association for Computing Machinery},
  \bibinfo{address}{New York, NY, USA}, \bibinfo{pages}{3739–3748}.
\newblock
\showISBNx{9781450331456}
\urldef\tempurl%
\url{https://doi.org/10.1145/2702123.2702529}
\showDOI{\tempurl}


\bibitem[Lapidot-Lefler and Barak(2012)]%
        {lapidot2012effects}
\bibfield{author}{\bibinfo{person}{Noam Lapidot-Lefler} {and}
  \bibinfo{person}{Azy Barak}.} \bibinfo{year}{2012}\natexlab{}.
\newblock \showarticletitle{Effects of anonymity, invisibility, and lack of
  eye-contact on toxic online disinhibition}.
\newblock \bibinfo{journal}{\emph{Computers in human behavior}}
  \bibinfo{volume}{28}, \bibinfo{number}{2} (\bibinfo{year}{2012}),
  \bibinfo{pages}{434--443}.
\newblock


\bibitem[Lapolla(2020)]%
        {lapolla2020tackling}
\bibfield{author}{\bibinfo{person}{Matthew Lapolla}.}
  \bibinfo{year}{2020}\natexlab{}.
\newblock \showarticletitle{Tackling Toxicity: Identifying and Addressing Toxic
  Behavior in Online Video Games}.
\newblock \bibinfo{journal}{\emph{Seton Hall University Dissertations and
  Theses (ETDs)}}  \bibinfo{volume}{2798} (\bibinfo{year}{2020}),
  \bibinfo{numpages}{70}~pages.
\newblock


\bibitem[Lechner(2010)]%
        {lechner2011}
\bibfield{author}{\bibinfo{person}{Michael Lechner}.}
  \bibinfo{year}{2010}\natexlab{}.
\newblock \showarticletitle{The Estimation of Causal Effects by
  Difference-in-Difference Methods}.
\newblock \bibinfo{journal}{\emph{Foundations and Trends{\textregistered} in
  Econometrics}} \bibinfo{volume}{4}, \bibinfo{number}{3}
  (\bibinfo{year}{2010}), \bibinfo{pages}{165--224}.
\newblock
\urldef\tempurl%
\url{https://doi.org/10.1561/0800000014}
\showDOI{\tempurl}


\bibitem[Lee et~al\mbox{.}(2019)]%
        {lee2019disruptive}
\bibfield{author}{\bibinfo{person}{Sung~Je Lee}, \bibinfo{person}{Eui~Jun
  Jeong}, {and} \bibinfo{person}{Joon~Hyun Jeon}.}
  \bibinfo{year}{2019}\natexlab{}.
\newblock \showarticletitle{Disruptive behaviors in online games: effects of
  moral positioning, competitive motivation, and aggression in “League of
  Legends”}.
\newblock \bibinfo{journal}{\emph{Social Behavior and Personality: an
  international journal}} \bibinfo{volume}{47}, \bibinfo{number}{2}
  (\bibinfo{year}{2019}), \bibinfo{pages}{1--9}.
\newblock


\bibitem[Levkovitz(2023)]%
        {Contentm60:online}
\bibfield{author}{\bibinfo{person}{Zohar Levkovitz}.}
  \bibinfo{year}{2023}\natexlab{}.
\newblock \bibinfo{title}{Content moderators alone can't clean up our toxic
  internet}.
\newblock
  \bibinfo{howpublished}{\url{https://www.fastcompany.com/90515733/content-moderators-alone-cant-clean-up-our-toxic-internet}}.
\newblock
\newblock
\shownote{(Accessed on 04/20/2023)}.


\bibitem[Link et~al\mbox{.}(2016)]%
        {link2016human}
\bibfield{author}{\bibinfo{person}{Daniel Link}, \bibinfo{person}{Bernd
  Hellingrath}, {and} \bibinfo{person}{Jie Ling}.}
  \bibinfo{year}{2016}\natexlab{}.
\newblock \showarticletitle{A Human-is-the-Loop Approach for Semi-Automated
  Content Moderation}. In \bibinfo{booktitle}{\emph{Proceedings of the ISCRAM
  2016 Conference}}. \bibinfo{publisher}{ISCRAM}, \bibinfo{address}{Rio de
  Janeiro, Brazil}.
\newblock


\bibitem[Liu and Agur(2023)]%
        {liu2023after}
\bibfield{author}{\bibinfo{person}{Yansheng Liu} {and} \bibinfo{person}{Colin
  Agur}.} \bibinfo{year}{2023}\natexlab{}.
\newblock \showarticletitle{“After All, They Don’t Know Me” Exploring the
  psychological mechanisms of toxic behavior in online games}.
\newblock \bibinfo{journal}{\emph{Games and Culture}} \bibinfo{volume}{18},
  \bibinfo{number}{5} (\bibinfo{year}{2023}), \bibinfo{pages}{598--621}.
\newblock


\bibitem[Lykouris and Weng(2024)]%
        {lykouris2024learning}
\bibfield{author}{\bibinfo{person}{Thodoris Lykouris} {and}
  \bibinfo{person}{Wentao Weng}.} \bibinfo{year}{2024}\natexlab{}.
\newblock \showarticletitle{Learning to defer in content moderation: The
  human-ai interplay}.
\newblock \bibinfo{journal}{\emph{arXiv preprint arXiv:2402.12237}}
  (\bibinfo{year}{2024}).
\newblock


\bibitem[Lyu et~al\mbox{.}(2024)]%
        {lyu2024got}
\bibfield{author}{\bibinfo{person}{Yao Lyu}, \bibinfo{person}{Jie Cai},
  \bibinfo{person}{Anisa Callis}, \bibinfo{person}{Kelley Cotter}, {and}
  \bibinfo{person}{John~M Carroll}.} \bibinfo{year}{2024}\natexlab{}.
\newblock \showarticletitle{"I Got Flagged for Supposed Bullying, Even Though
  It Was in Response to Someone Harassing Me About My Disability.": A Study of
  Blind TikTokers' Content Moderation Experiences}.
\newblock \bibinfo{journal}{\emph{arXiv e-prints}}  \bibinfo{volume}{1}
  (\bibinfo{year}{2024}), \bibinfo{pages}{arXiv--2401}.
\newblock


\bibitem[Ma et~al\mbox{.}(2023)]%
        {ma2023transparency}
\bibfield{author}{\bibinfo{person}{Renkai Ma}, \bibinfo{person}{Yao Li}, {and}
  \bibinfo{person}{Yubo Kou}.} \bibinfo{year}{2023}\natexlab{}.
\newblock \showarticletitle{Transparency, Fairness, and Coping: How Players
  Experience Moderation in Multiplayer Online Games}. In
  \bibinfo{booktitle}{\emph{Proceedings of the 2023 CHI Conference on Human
  Factors in Computing Systems}}. \bibinfo{publisher}{Association for Computing
  Machinery}, \bibinfo{address}{Hamburg, Germany}, \bibinfo{pages}{1--21}.
\newblock


\bibitem[Migliore(2021)]%
        {migliore2021esports}
\bibfield{author}{\bibinfo{person}{Lindsey Migliore}.}
  \bibinfo{year}{2021}\natexlab{}.
\newblock \showarticletitle{What is esports? The past, present, and future of
  competitive gaming}.
\newblock In \bibinfo{booktitle}{\emph{Handbook of Esports Medicine: Clinical
  Aspects of Competitive Video Gaming}}. \bibinfo{publisher}{Springer},
  \bibinfo{pages}{1--16}.
\newblock


\bibitem[Nederhof(1985)]%
        {nederhof1985methods}
\bibfield{author}{\bibinfo{person}{Anton~J Nederhof}.}
  \bibinfo{year}{1985}\natexlab{}.
\newblock \showarticletitle{Methods of coping with social desirability bias: A
  review}.
\newblock \bibinfo{journal}{\emph{European journal of social psychology}}
  \bibinfo{volume}{15}, \bibinfo{number}{3} (\bibinfo{year}{1985}),
  \bibinfo{pages}{263--280}.
\newblock


\bibitem[Neto et~al\mbox{.}(2017)]%
        {neto2017studying}
\bibfield{author}{\bibinfo{person}{Joaquim~AM Neto}, \bibinfo{person}{Kazuki~M
  Yokoyama}, {and} \bibinfo{person}{Karin Becker}.}
  \bibinfo{year}{2017}\natexlab{}.
\newblock \showarticletitle{Studying toxic behavior influence and player chat
  in an online video game}. In \bibinfo{booktitle}{\emph{Proceedings of the
  international conference on web intelligence}}. \bibinfo{pages}{26--33}.
\newblock


\bibitem[Paschke et~al\mbox{.}(2021)]%
        {paschke2021adolescent}
\bibfield{author}{\bibinfo{person}{Kerstin Paschke},
  \bibinfo{person}{Maria~Isabella Austermann}, \bibinfo{person}{Kathrin
  Simon-Kutscher}, {and} \bibinfo{person}{Rainer Thomasius}.}
  \bibinfo{year}{2021}\natexlab{}.
\newblock \showarticletitle{Adolescent gaming and social media usage before and
  during the COVID-19 pandemic}.
\newblock \bibinfo{journal}{\emph{Sucht}}  \bibinfo{volume}{67}
  (\bibinfo{year}{2021}), \bibinfo{pages}{1664--2856}.
\newblock
Issue 1.


\bibitem[Patel et~al\mbox{.}(2021)]%
        {patel2021user}
\bibfield{author}{\bibinfo{person}{Aashka Patel}, \bibinfo{person}{Christine~L
  Cook}, {and} \bibinfo{person}{Donghee~Yvette Wohn}.}
  \bibinfo{year}{2021}\natexlab{}.
\newblock \showarticletitle{User Opinions on Effective Strategies Against
  Social Media Toxicity}. In \bibinfo{booktitle}{\emph{Proceedings of the
  Annual Hawaii International Conference on System Sciences}}.
  \bibinfo{publisher}{IEEE}, \bibinfo{address}{Hawaii},
  \bibinfo{pages}{3005--3014}.
\newblock


\bibitem[Pratt et~al\mbox{.}(2017)]%
        {pratt2017empirical}
\bibfield{author}{\bibinfo{person}{Travis~C Pratt}, \bibinfo{person}{Francis~T
  Cullen}, \bibinfo{person}{Kristie~R Blevins}, \bibinfo{person}{Leah~E
  Daigle}, {and} \bibinfo{person}{Tamara~D Madensen}.}
  \bibinfo{year}{2017}\natexlab{}.
\newblock \showarticletitle{The empirical status of deterrence theory: A
  meta-analysis}.
\newblock In \bibinfo{booktitle}{\emph{Taking stock}}.
  \bibinfo{publisher}{Routledge}, \bibinfo{address}{London, UK},
  \bibinfo{pages}{367--395}.
\newblock


\bibitem[Raith et~al\mbox{.}(2021)]%
        {raith2021massively}
\bibfield{author}{\bibinfo{person}{Lisa Raith}, \bibinfo{person}{Julie
  Bignill}, \bibinfo{person}{Vasileios Stavropoulos}, \bibinfo{person}{Prudence
  Millear}, \bibinfo{person}{Andrew Allen}, \bibinfo{person}{Helen~M Stallman},
  \bibinfo{person}{Jonathan Mason}, \bibinfo{person}{Tamara De~Regt},
  \bibinfo{person}{Andrew Wood}, {and} \bibinfo{person}{Lee Kannis-Dymand}.}
  \bibinfo{year}{2021}\natexlab{}.
\newblock \showarticletitle{Massively multiplayer online games and well-being:
  A systematic literature review}.
\newblock \bibinfo{journal}{\emph{Frontiers in Psychology}}
  \bibinfo{volume}{12} (\bibinfo{year}{2021}), \bibinfo{pages}{698799}.
\newblock


\bibitem[Rieder and Skop(2021)]%
        {rieder2021fabrics}
\bibfield{author}{\bibinfo{person}{Bernhard Rieder} {and}
  \bibinfo{person}{Yarden Skop}.} \bibinfo{year}{2021}\natexlab{}.
\newblock \showarticletitle{The fabrics of machine moderation: Studying the
  technical, normative, and organizational structure of Perspective API}.
\newblock \bibinfo{journal}{\emph{Big Data \& Society}} \bibinfo{volume}{8},
  \bibinfo{number}{2} (\bibinfo{year}{2021}),
  \bibinfo{pages}{20539517211046181}.
\newblock


\bibitem[RockedBrush(2024)]%
        {TopVideo36:online}
\bibfield{author}{\bibinfo{person}{RockedBrush}.}
  \bibinfo{year}{2024}\natexlab{}.
\newblock \bibinfo{title}{Top Video Game Genres in 2024: Revenue, Statistics}.
\newblock
  \bibinfo{howpublished}{\url{https://rocketbrush.com/blog/most-popular-video-game-genres-in-2024-revenue-statistics-genres-overview}}.
\newblock
\newblock
\shownote{(Accessed on 09/23/2024)}.


\bibitem[Rosenbaum and Rubin(1983)]%
        {rosenbaumRubin1983}
\bibfield{author}{\bibinfo{person}{Paul~R Rosenbaum} {and}
  \bibinfo{person}{Donald~B Rubin}.} \bibinfo{year}{1983}\natexlab{}.
\newblock \showarticletitle{The central role of the propensity score in
  observational studies for causal effects}.
\newblock \bibinfo{journal}{\emph{Biometrika}} \bibinfo{volume}{70},
  \bibinfo{number}{1} (\bibinfo{year}{1983}), \bibinfo{pages}{41--55}.
\newblock


\bibitem[Saarinen(2017)]%
        {saarinen2017toxic}
\bibfield{author}{\bibinfo{person}{Teemu Saarinen}.}
  \bibinfo{year}{2017}\natexlab{}.
\newblock \emph{\bibinfo{title}{Toxic behavior in online games}}.
\newblock \bibinfo{thesistype}{Master's\ thesis}. \bibinfo{school}{T.
  Saarinen}.
\newblock


\bibitem[Sant’Anna and Zhao(2020)]%
        {sant2020doubly}
\bibfield{author}{\bibinfo{person}{Pedro~HC Sant’Anna} {and}
  \bibinfo{person}{Jun Zhao}.} \bibinfo{year}{2020}\natexlab{}.
\newblock \showarticletitle{Doubly robust difference-in-differences
  estimators}.
\newblock \bibinfo{journal}{\emph{Journal of Econometrics}}
  \bibinfo{volume}{219}, \bibinfo{number}{1} (\bibinfo{year}{2020}),
  \bibinfo{pages}{101--122}.
\newblock


\bibitem[Sch{\"o}lkopf(2022)]%
        {scholkopf2022causality}
\bibfield{author}{\bibinfo{person}{Bernhard Sch{\"o}lkopf}.}
  \bibinfo{year}{2022}\natexlab{}.
\newblock \showarticletitle{Causality for machine learning}.
\newblock In \bibinfo{booktitle}{\emph{Probabilistic and causal inference: The
  works of Judea Pearl}}. \bibinfo{pages}{765--804}.
\newblock


\bibitem[Shiba and Kawahara(2021)]%
        {shiba2021using}
\bibfield{author}{\bibinfo{person}{Koichiro Shiba} {and}
  \bibinfo{person}{Takuya Kawahara}.} \bibinfo{year}{2021}\natexlab{}.
\newblock \showarticletitle{Using propensity scores for causal inference:
  pitfalls and tips}.
\newblock \bibinfo{journal}{\emph{Journal of epidemiology}}
  \bibinfo{volume}{31}, \bibinfo{number}{8} (\bibinfo{year}{2021}),
  \bibinfo{pages}{457--463}.
\newblock


\bibitem[Shim et~al\mbox{.}(2011)]%
        {shim2011exploratory}
\bibfield{author}{\bibinfo{person}{Kyong~Jin Shim}, \bibinfo{person}{Kuo-Wei
  Hsu}, \bibinfo{person}{Samarth Damania}, \bibinfo{person}{Colin DeLong},
  {and} \bibinfo{person}{Jaideep Srivastava}.} \bibinfo{year}{2011}\natexlab{}.
\newblock \showarticletitle{An exploratory study of player and team performance
  in multiplayer first-person-shooter games}. In \bibinfo{booktitle}{\emph{2011
  IEEE Third International Conference on Privacy, Security, Risk and Trust and
  2011 IEEE Third International Conference on Social Computing}}. IEEE,
  \bibinfo{pages}{617--620}.
\newblock


\bibitem[Srikanth et~al\mbox{.}(2021)]%
        {srikanth2021dynamic}
\bibfield{author}{\bibinfo{person}{Maya Srikanth}, \bibinfo{person}{Anqi Liu},
  \bibinfo{person}{Nicholas Adams-Cohen}, \bibinfo{person}{Jian Cao},
  \bibinfo{person}{R~Michael Alvarez}, {and} \bibinfo{person}{Anima
  Anandkumar}.} \bibinfo{year}{2021}\natexlab{}.
\newblock \showarticletitle{Dynamic social media monitoring for fast-evolving
  online discussions}. In \bibinfo{booktitle}{\emph{Proceedings of the 27th ACM
  SIGKDD Conference on Knowledge Discovery \& Data Mining}}.
  \bibinfo{publisher}{Association for Computing Machinery},
  \bibinfo{address}{Singapore}, \bibinfo{pages}{3576--3584}.
\newblock


\bibitem[Srinivasan et~al\mbox{.}(2019)]%
        {srinivasan2019content}
\bibfield{author}{\bibinfo{person}{Kumar~Bhargav Srinivasan},
  \bibinfo{person}{Cristian Danescu-Niculescu-Mizil}, \bibinfo{person}{Lillian
  Lee}, {and} \bibinfo{person}{Chenhao Tan}.} \bibinfo{year}{2019}\natexlab{}.
\newblock \showarticletitle{Content removal as a moderation strategy:
  Compliance and other outcomes in the changemyview community}.
\newblock \bibinfo{journal}{\emph{Proceedings of the ACM on Human-Computer
  Interaction}} \bibinfo{volume}{3}, \bibinfo{number}{CSCW}
  (\bibinfo{year}{2019}), \bibinfo{pages}{1--21}.
\newblock


\bibitem[Steinkuehler(2023)]%
        {steinkuehler2023games}
\bibfield{author}{\bibinfo{person}{Constance Steinkuehler}.}
  \bibinfo{year}{2023}\natexlab{}.
\newblock \showarticletitle{Games as social platforms}.
\newblock \bibinfo{journal}{\emph{ACM Games: Research and Practice}}
  \bibinfo{volume}{1}, \bibinfo{number}{1} (\bibinfo{year}{2023}),
  \bibinfo{pages}{1--2}.
\newblock


\bibitem[Stoop et~al\mbox{.}(2019)]%
        {stoop2019detecting}
\bibfield{author}{\bibinfo{person}{Wessel Stoop}, \bibinfo{person}{Florian
  Kunneman}, \bibinfo{person}{Antal van~den Bosch}, {and} \bibinfo{person}{Ben
  Miller}.} \bibinfo{year}{2019}\natexlab{}.
\newblock \showarticletitle{Detecting harassment in real-time as conversations
  develop}. In \bibinfo{booktitle}{\emph{Proceedings of the Third Workshop on
  Abusive Language Online}}. \bibinfo{publisher}{Association for Computational
  Linguistics}, \bibinfo{address}{Florence, Italy}, \bibinfo{pages}{19--24}.
\newblock


\bibitem[Trepte et~al\mbox{.}(2012)]%
        {trepte2012social}
\bibfield{author}{\bibinfo{person}{Sabine Trepte}, \bibinfo{person}{Leonard
  Reinecke}, {and} \bibinfo{person}{Keno Juechems}.}
  \bibinfo{year}{2012}\natexlab{}.
\newblock \showarticletitle{The social side of gaming: How playing online
  computer games creates online and offline social support}.
\newblock \bibinfo{journal}{\emph{Computers in Human behavior}}
  \bibinfo{volume}{28}, \bibinfo{number}{3} (\bibinfo{year}{2012}),
  \bibinfo{pages}{832--839}.
\newblock


\bibitem[Tucker et~al\mbox{.}(2023)]%
        {tucker2023bandits}
\bibfield{author}{\bibinfo{person}{Aaron~D Tucker}, \bibinfo{person}{Caleb
  Biddulph}, \bibinfo{person}{Claire Wang}, {and} \bibinfo{person}{Thorsten
  Joachims}.} \bibinfo{year}{2023}\natexlab{}.
\newblock \showarticletitle{Bandits with costly reward observations}. In
  \bibinfo{booktitle}{\emph{Uncertainty in Artificial Intelligence}}. PMLR,
  \bibinfo{pages}{2147--2156}.
\newblock


\bibitem[Vaccaro et~al\mbox{.}(2018)]%
        {vaccaro2018illusion}
\bibfield{author}{\bibinfo{person}{Kristen Vaccaro}, \bibinfo{person}{Dylan
  Huang}, \bibinfo{person}{Motahhare Eslami}, \bibinfo{person}{Christian
  Sandvig}, \bibinfo{person}{Kevin Hamilton}, {and} \bibinfo{person}{Karrie
  Karahalios}.} \bibinfo{year}{2018}\natexlab{}.
\newblock \showarticletitle{The illusion of control: Placebo effects of control
  settings}. In \bibinfo{booktitle}{\emph{Proceedings of the 2018 CHI
  Conference on Human Factors in Computing Systems}}. \bibinfo{pages}{1--13}.
\newblock


\bibitem[Weld et~al\mbox{.}(2021)]%
        {weld2021conda}
\bibfield{author}{\bibinfo{person}{Henry Weld}, \bibinfo{person}{Guanghao
  Huang}, \bibinfo{person}{Jean Lee}, \bibinfo{person}{Tongshu Zhang},
  \bibinfo{person}{Kunze Wang}, \bibinfo{person}{Xinghong Guo},
  \bibinfo{person}{Siqu Long}, \bibinfo{person}{Josiah Poon}, {and}
  \bibinfo{person}{Soyeon~Caren Han}.} \bibinfo{year}{2021}\natexlab{}.
\newblock \showarticletitle{Conda: a contextual dual-annotated dataset for
  in-game toxicity understanding and detection}. In
  \bibinfo{booktitle}{\emph{Findings of the Association for Computational
  Linguistics: ACL-IJCNLP}}, Vol.~\bibinfo{volume}{abs/2106.06213}.
  \bibinfo{publisher}{Association for Computational Linguistics},
  \bibinfo{address}{Bangkok, Thailand}, \bibinfo{pages}{2406--2416}.
\newblock
\urldef\tempurl%
\url{https://api.semanticscholar.org/CorpusID:235417027}
\showURL{%
\tempurl}


\bibitem[Wen(2019)]%
        {Wen2019-WENDDI}
\bibfield{author}{\bibinfo{person}{Wen Wen}.} \bibinfo{year}{2019}\natexlab{}.
\newblock \showarticletitle{Does Delay in Feedback Diminish Sense of Agency? A
  Review}.
\newblock \bibinfo{journal}{\emph{Consciousness and Cognition}}
  \bibinfo{volume}{73} (\bibinfo{year}{2019}), \bibinfo{pages}{102759}.
\newblock
\urldef\tempurl%
\url{https://doi.org/10.1016/j.concog.2019.05.007}
\showDOI{\tempurl}


\bibitem[West et~al\mbox{.}(2000)]%
        {west2000causal}
\bibfield{author}{\bibinfo{person}{Stephen~G West}, \bibinfo{person}{Jeremy~C
  Biesanz}, {and} \bibinfo{person}{Steven~C Pitts}.}
  \bibinfo{year}{2000}\natexlab{}.
\newblock \showarticletitle{Causal inference and generalization in field
  settings: Experimental and quasi-experimental designs.}
\newblock  (\bibinfo{year}{2000}).
\newblock


\bibitem[Wijkstra et~al\mbox{.}(2023)]%
        {wijkstra2023help}
\bibfield{author}{\bibinfo{person}{Michel Wijkstra}, \bibinfo{person}{Katja
  Rogers}, \bibinfo{person}{Regan~L Mandryk}, \bibinfo{person}{Remco~C
  Veltkamp}, {and} \bibinfo{person}{Julian Frommel}.}
  \bibinfo{year}{2023}\natexlab{}.
\newblock \showarticletitle{Help, My Game Is Toxic! First Insights from a
  Systematic Literature Review on Intervention Systems for Toxic Behaviors in
  Online Video Games}. In \bibinfo{booktitle}{\emph{Companion Proceedings of
  the Annual Symposium on Computer-Human Interaction in Play}}.
  \bibinfo{publisher}{Association for Computing Machinery},
  \bibinfo{address}{New York, NY, USA}, \bibinfo{pages}{3--9}.
\newblock
\showISBNx{9798400700293}


\bibitem[Wiki(2021)]%
        {KilltoDe71:online}
\bibfield{author}{\bibinfo{person}{LoL Wiki}.} \bibinfo{year}{2021}\natexlab{}.
\newblock \bibinfo{title}{Kill to Death Ratio -League of Legends Wiki -
  Fandom}.
\newblock
  \bibinfo{howpublished}{\url{https://leagueoflegends.fandom.com/wiki/Kill_to_Death_Ratio\#:~:text=The
  Kill to Death ratio, other players in the game.}}.
\newblock
\newblock
\shownote{(Accessed on 10/09/2024)}.


\bibitem[Wojcik et~al\mbox{.}(2022)]%
        {wojck2022birdwatch}
\bibfield{author}{\bibinfo{person}{Stefan Wojcik}, \bibinfo{person}{Sophie
  Hilgard}, \bibinfo{person}{Nick Judd}, \bibinfo{person}{Delia Mocanu},
  \bibinfo{person}{Stephen Ragain}, \bibinfo{person}{M.~B.~Fallin Hunzaker},
  \bibinfo{person}{Keith Coleman}, {and} \bibinfo{person}{Jay Baxter}.}
  \bibinfo{year}{2022}\natexlab{}.
\newblock \bibinfo{title}{Birdwatch: Crowd Wisdom and Bridging Algorithms can
  Inform Understanding and Reduce the Spread of Misinformation}.
\newblock
\newblock
\showeprint[arxiv]{2210.15723}~[cs.SI]


\bibitem[Zhang and Naidu(2024)]%
        {zhang2024sido}
\bibfield{author}{\bibinfo{person}{Amy~X Zhang} {and} \bibinfo{person}{Parth
  Naidu}.} \bibinfo{year}{2024}\natexlab{}.
\newblock \showarticletitle{The SIDO Performance Model for League of Legends}.
\newblock \bibinfo{journal}{\emph{arXiv preprint arXiv:2403.04873}}
  (\bibinfo{year}{2024}).
\newblock


\bibitem[Zhang et~al\mbox{.}(2023)]%
        {zhang2023china}
\bibfield{author}{\bibinfo{person}{Xiaohui Zhang}, \bibinfo{person}{Zaiyan
  Wei}, \bibinfo{person}{Qianzhou Du}, {and} \bibinfo{person}{Zhongju Zhang}.}
  \bibinfo{year}{2023}\natexlab{}.
\newblock \showarticletitle{Social Media Moderation and Content Generation:
  Evidence from User Bans}.
\newblock \bibinfo{journal}{\emph{SSRN}} \bibinfo{volume}{1},
  \bibinfo{number}{1} (\bibinfo{year}{2023}), \bibinfo{numpages}{69}~pages.
\newblock
\newblock
\shownote{Available at SSRN: https://ssrn.com/abstract=4089011 or
  http://dx.doi.org/10.2139/ssrn.4089011}.


\end{thebibliography}

\newpage

\renewcommand{\thesection}{S\arabic{section}}
\setcounter{section}{0}
\renewcommand{\thefigure}{S\arabic{figure}}
\setcounter{figure}{0}
\renewcommand{\thetable}{S\arabic{table}}
\setcounter{table}{0}

\section{Supplementary Information}

\subsection{Frequency of Individual Moderation Actions}
\label{apx:mod-action-freq}

In Table \ref{tab:reason_action} we report the frequencies of different moderation actions applied within the context of different types of
offenses.

\begin{table}[!htp]
\small
\centering
  \begin{adjustbox}{width=\textwidth}

\small 
\begin{tabular}{llr}
\hline
\textbf{Moderation Reason} & \textbf{Actions Taken} & \textbf{Ratio} \\
\hline
Cheater & Remove From Leaderboards & 97.37\% \\
Cheater & Ranking Service, Remove From Leaderboards & 2.63\% \\

\midrule

Offensive text chat & Feature Flag & 0.09\% \\
Offensive text chat & Penalty Notice & 0.05\% \\
Offensive text chat & Penalty Notice, Feature Flag & 99.55\% \\
Offensive text chat & Warning Notice, Feature Flag & 0.31\% \\

\midrule

Offensive user identification & Delete Profile, Rename User, & \\
& Limit Allowed Renames & 0.01\% \\
Offensive user identification & Delete Profile, Rename User, & \\
& Limit Allowed Renames, Remove Clantag & 0.09\% \\
Offensive user identification & Delete Profile, Rename User, & \\
& Limit Allowed Renames, Remove Clantag & 0.14\% \\
Offensive user identification & Limit Allowed Renames, & \\
& Update Clantag, Penalty Notice, Feature Flag & 0.03\% \\
Offensive user identification & Warning Notice & 0.19\% \\
Offensive user identification & Rename User, Limit Allowed Renames & 1.15\% \\
Offensive user identification & Rename User, Limit Allowed Renames, & \\
& Penalty Notice, Feature Flag & 0.10\% \\
Offensive user identification & Rename User, Limit Allowed Renames, & \\
& Update Clantag, Penalty Notice & 5.49\% \\
Offensive user identification & Rename User, Limit Allowed Renames, & \\
& Update Clantag, Penalty Notice, Feature Flag & 92.81\% \\

\midrule

Offensive voice chat & Feature Flag & 26.75\% \\
Offensive voice chat & Feature Flag, Feature Flag, Penalty Notice & 41.78\% \\
Offensive voice chat & Feature Flag, Penalty Notice & 0.01\% \\
Offensive voice chat & Warning Notice & 0.01\% \\
Offensive voice chat & Warning Notice, Feature Flag & 31.46\% \\
\hline
\end{tabular}
\end{adjustbox}
\caption{Moderation actions and their relative frequency in our dataset. In many cases, multiple actions are taken simultaneously to moderate the player.}
\label{tab:reason_action}
\end{table}

\newpage

\subsection{Estimating CATE with and without the propensity score}

\begin{figure}[ht]
    \centering
    \begin{subfigure}[b]{0.4\linewidth}
        \includegraphics[width=\linewidth]{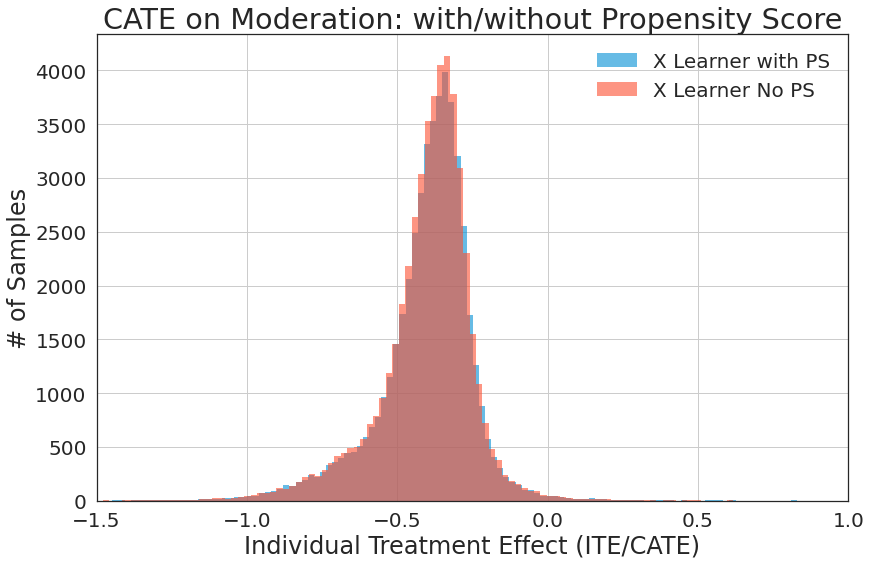}
        \caption{X Learner with/without Pscore}
    \end{subfigure}
    \hspace{1cm} 
    \begin{subfigure}[b]{0.4\linewidth}
        \includegraphics[width=\linewidth]{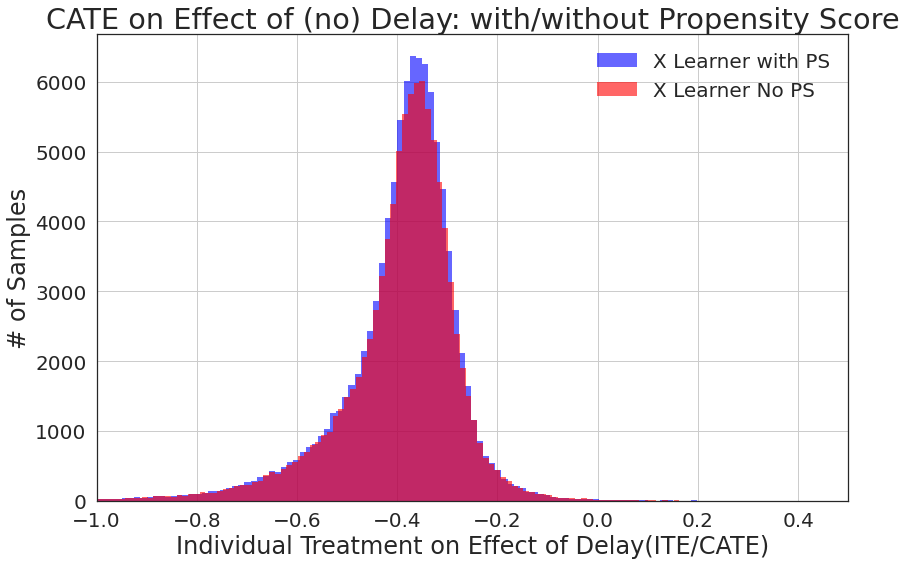}
        \caption{X Learner with/without Pscore}
    \end{subfigure}
    \caption{Robustness Checking: Comparing Learners with/without Pscore.}
    \Description{Robustness Checking: Comparing Learners with/without Pscore.}
    \label{fig:x-Learner-propensity}
\end{figure}

Fig \ref{fig:x-Learner-propensity} shows the impact of propensity score correction on the CATE estimated under X-Learner. We note the DR Learner we us for the main results, always uses propensity score, hence we plot the CATE estimates to the X-Learner.

\subsection{Data Pre-Processing}
\label{apx:data_preprocessing}

Our analysis focused on users moderated for cheating, offensive user IDs, or offensive behavior in text or voice chat from 2023-02-01 to 2023-04-19. We considered only the first moderation instance per user. In cases with multiple moderations, the longest delay before moderation was used to ensure a pessimistic bound on the effect of a shorter moderation delay. 

\begin{itemize}
 \item \textbf{Gather Report Data:} Collect data from three different types of player reports (cheating, offensive text chat, and offensive user identification) over a specified date range.
 \item \textbf{Process Moderation Data:} Combine the reports with data on moderation actions, categorizing each report based on the type of moderation action taken (if any).

 \item \textbf{Aggregate Data:} Aggregate the combined data by calculating number of reports by each moderation date, reasons, and actions, for each moderated players, calculating the time delay between each report and the corresponding moderation action. Join the gameplay (number of matches by player-date) records to the report records. 
 \item \textbf{Refine for Minimum Lag:} Filter the data to keep only those records where the delay between the report and the moderation action is the minimum for each reported player, based on the moderation date and reason. Ensure each player has only one record left to eliminate duplicates. 
 \item \textbf{Create Datasets for effect of moderation:} Label players by lag of reports, separate players in treatment group (lag $=$ 0) and control group (lag $>=$ 7). Calculate average participation and report rate in the 7-day window before and after receiving reports. 
  \item \textbf{Create Datasets for effect of delay:} Label players by lag of reports, separate players in treatment group (lag $<=$ 3) and control group (lag $>=$ 7). Calculate average participation and report rate in the 7-day window before receiving reports and 7-day window after receiving moderation. 
\end{itemize}

\subsection{Model Selection and Modeling Robustness Evaluation}
\label{apx:model_selection}

\begin{figure}[ht]
    \centering
    \begin{subfigure}[b]{0.4\linewidth}
        \includegraphics[width=\linewidth]{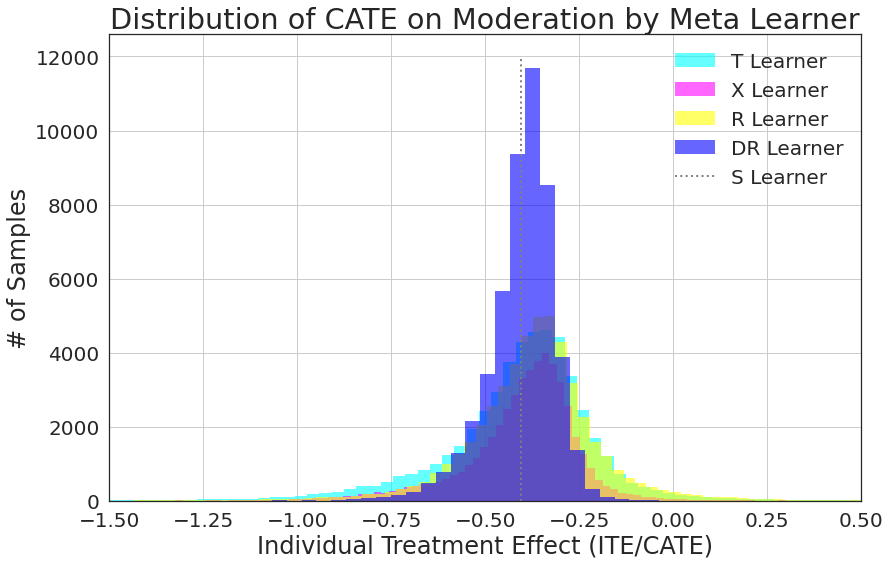}
        \caption{Distribution of  Conditional Average Treatment Effect (CATE) by Meta-Learner for impact of moderation vs no moderation.}
    \end{subfigure}
    \hspace{1cm} 
    \begin{subfigure}[b]{0.4\linewidth}
        \includegraphics[width=\linewidth]{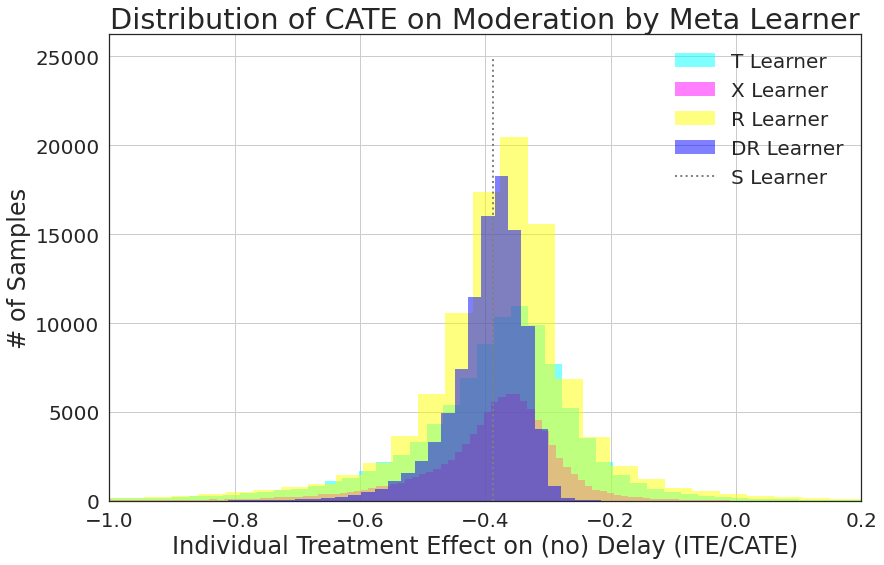}
        \caption{Distribution of Conditional Average Treatment Effect (CATE) by Meta-Learner for impact of immediate vs delayed moderation.}
    \end{subfigure}
    \caption{Robustness Checking: Comparing different Meta-Learners.} 
    \Description{Robustness Checking: Comparing different Meta-Learners.}
    \label{fig:robust}
\end{figure}

\paragraph{\textbf{Selection of Meta-Learner}:}
In Figure~\ref{fig:robust}, we display the CATE distribution of multiple Learners, including T-Learner, S-Learner, X-Learner, R-Learner, and Doubly Robust Learner. The two graphs indicate that all the Learners give relatively close results in the range of individual treatment effects (x-axis), providing evidence of the robustness of the result. In particular, the mean and mode of the estimates are close and significantly different from 0. Among all the Learners, DR Learner gives the highest concentration among all Learners, providing the most consistent estimates. Therefore, we proceed to use the DR Learner for our main results.

\paragraph{\textbf{Machine Learning Estimator Selection under Doubly Robust (DR) Meta-Learner}:}

\begin{figure}
    \centering
    \includegraphics[width=1.0\textwidth]{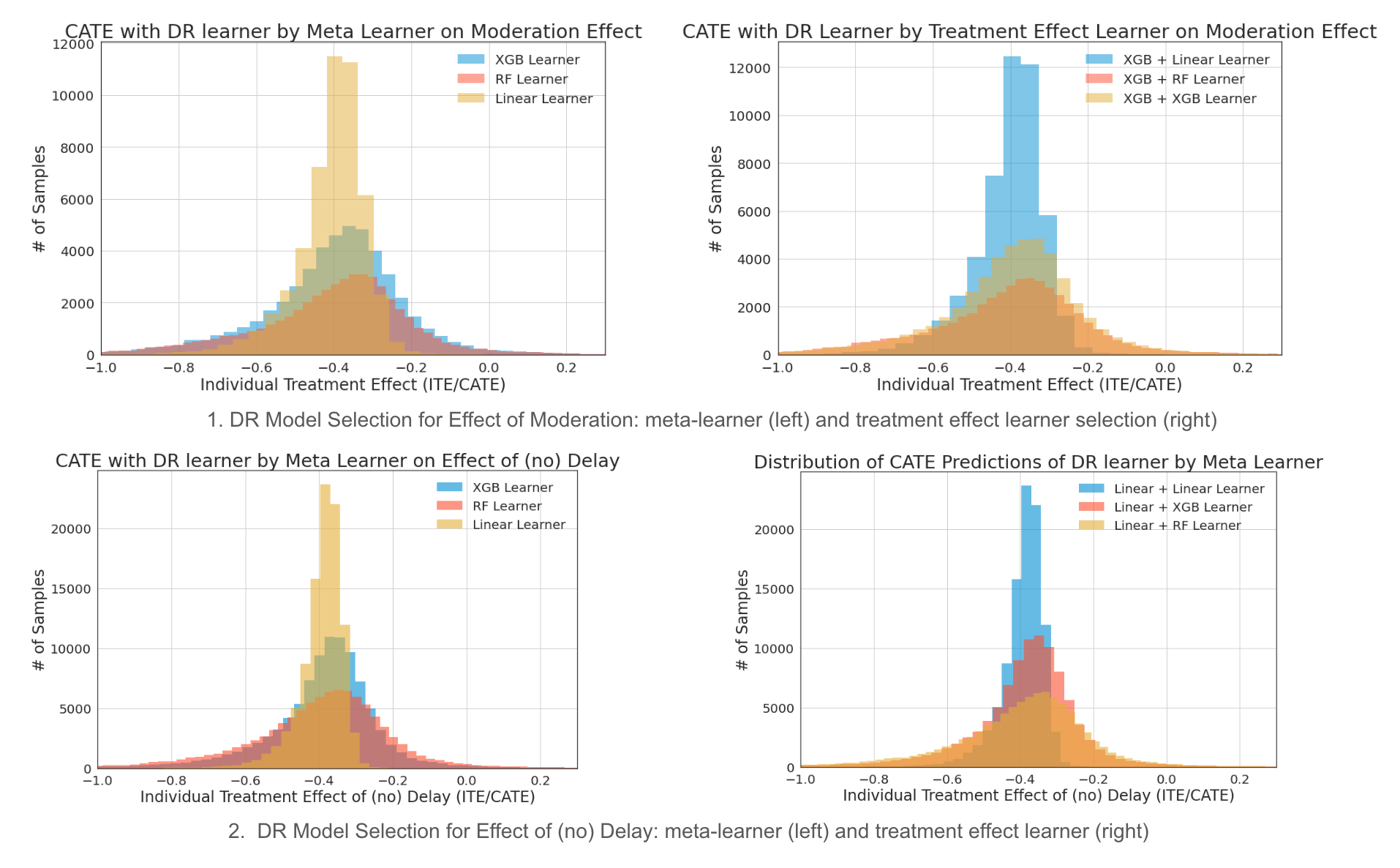}
    \caption{Based Model Selection (Machine Learning Estimator) Among DR Model Class. Top Left \& Bottom Left: selecting general Meta-Learner. Top Right \& Bottom Right: selecting Treatment Effect Learners.}
    \Description{Based Model Selection (Machine Learning Estimator) Among DR Model Class. Top Left \& Bottom Left: selecting general Meta-Learner. Top Right \& Bottom Right: selecting Treatment Effect Learners.}
    \label{fig:dr}
\end{figure}

We evaluate different base models (machine learning estimators) under the best DR meta-Learner. Figure~\ref{fig:dr} shows the distribution of CATE treatment effects using a doubly robust Meta-Learner while varying the type of the base Learner. In the first graph, we observe the performance of single models. The Random Forest Learner's distribution has the widest spread, suggesting a higher variance in its treatment effect estimates. The XGB Learner appears to have the narrowest and most symmetric distribution around zero, which suggests less variability and more conservative estimates of the treatment effect.

Aside from the base Learner used to estimate outcomes and treatment effects in both the control and treatment groups, we can also select a treatment-effect Learner. This model is used to estimate the treatment effects in the treatment group. Based on the best-performing base Learner (XGB in our case), we further vary the machine learning estimator used for the treatment effect Learner. The second graph shows combined Learners. Here, the XGB+Linear Learner's distribution is the narrowest, implying highly consistent treatment effect estimates. The XGB+RF Learner and XGB+XGB Learner have wider distributions, indicating less consistency in their predictions.

Similarly to effect of moderation, we also perform the same 2-stage Learner selection for the effect of delay. In the third graph for single models, the Linear Learner has the highest peak with the smallest spread, indicating the least variance in estimation. Conditional on the Linear base Learner, we further select linear Learner as the treatment effect estimator according to the fourth graph since it offers the most consistent treatment effect estimates

\begin{figure}[t]
    \centering
    \includegraphics[width=1.0\textwidth]{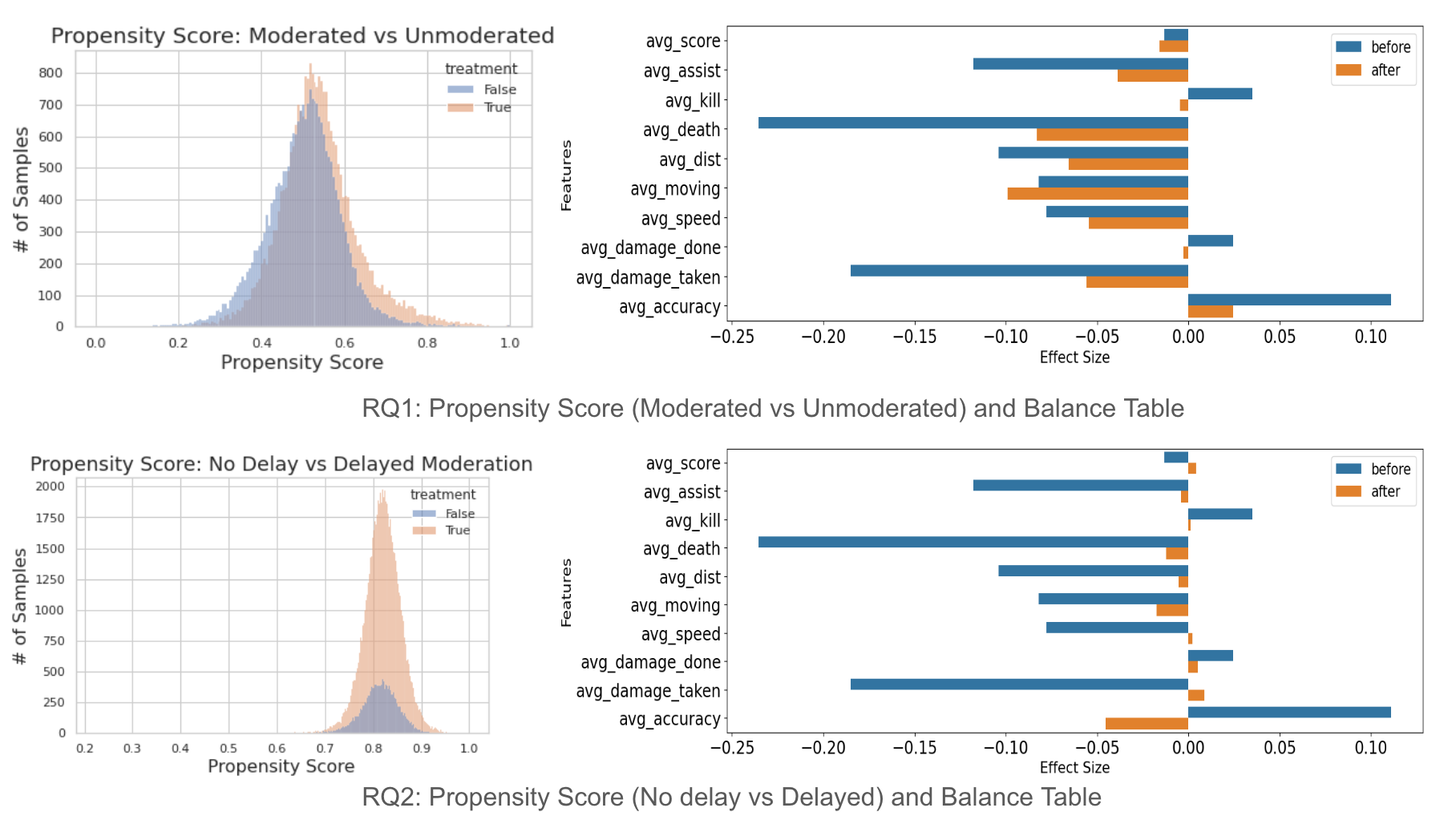}
    \caption{Propensity Score distribution and feature balance. Top Left \& Bottom Left: distribution Propensity Scores. Top Right \& Bottom Right: Standardized Mean differences across covariates before and after matching.}
    \Description{Propensity Score distribution and feature balance. Top Left \& Bottom Left: distribution Propensity Scores. Top Right \& Bottom Right: Standardized Mean differences across covariates before and after matching.}
    \label{fig:pscore}
\end{figure}

\paragraph{\textbf{Propensity Modeling}:} To correct for the potential systematic difference in participant assignment to treatment vs control, we calculate the propensity score (balancing score) and estimate its impact on outcome estimates. Figure~\ref{fig:pscore} gives the distribution of propensity score among treatment and control groups. The graphs indicate that in both propensity estimation, propensity score distribution for treatment and control group are in a similar range. The difference in peak is due to the treatment/control group size being more balanced in the data we use for the effect of moderation, where the number of players who are moderated and not are roughly the same. For the effect of delay, since the majority of the players are moderated within 3 days, we have fewer players in the control group. 

In both graphs, treatment and control groups overlap significantly, implying that for any given propensity score, there are individuals from both the treatment and control groups. 

Figure~\ref{fig:pscore} also shows the contribution of each feature to the propensity score after we do propensity score matching with KNN. The graph shows that most features become more balanced after matching with a smaller effect size.  This provides the evidence for the validity of the propensity score. The graphs provide reassurance that the propensity scores are capturing information on the likelihood of treatment.

In Table~\ref{tab:features} we summarize the gaming features we use to estimate the propensity score and display the mean value of each group.

\begin{table}[ht]
\centering
\begin{tabular}{lrrrr}

\hline
\textbf{Features (mean)}          & \textbf{ Unmoderated} & \textbf{ Moderated} & \textbf{Delayed Moderation} & \textbf{Immediate Moderation} \\ \hline
Score      & 2156.4   & 2137.5   & 2250.1   & 2257.6   \\
Assist      & 3.3      & 3.0      & 3.4      & 3.3      \\
Elimination        & 15.1     & 15.4     & 15.7     & 16.1     \\
Death       & 13.2     & 11.7     & 13.4     & 12.6     \\
Distance       & 42646.9  & 41099.2  & 43232.9  & 42612.5  \\
Moving       & 77.9     & 76.8     & 78.4     & 78.1     \\ 
Speed        & 135.7    & 132.7    & 137.3    & 136.6    \\ 
Damage done & 1619.5   & 1649.8   & 1680.4   & 1711.2   \\ 
Damage taken & 1447.5   & 1300.2   & 1472.7   & 1396.4   \\
Accuracy & 20.6     & 21.1     & 20.8     & 21.1 \\ \hline
\end{tabular}
\caption{Feature Balance Table.
}
\label{tab:features}
\end{table}

\subsection{Heterogeneity of Moderation Effects}
\label{apx:heterogeneity}

\begin{figure}[htbp]
    \caption{Correlations of conditional average treatment effect (CATE) for change in Report Rates and player performance characteristics.}
    \Description{Correlation of CATE for Report Rates and Player Performance Characteristics.}
    \label{fig:cate_on_reports}
    \includegraphics[width=.80\linewidth]{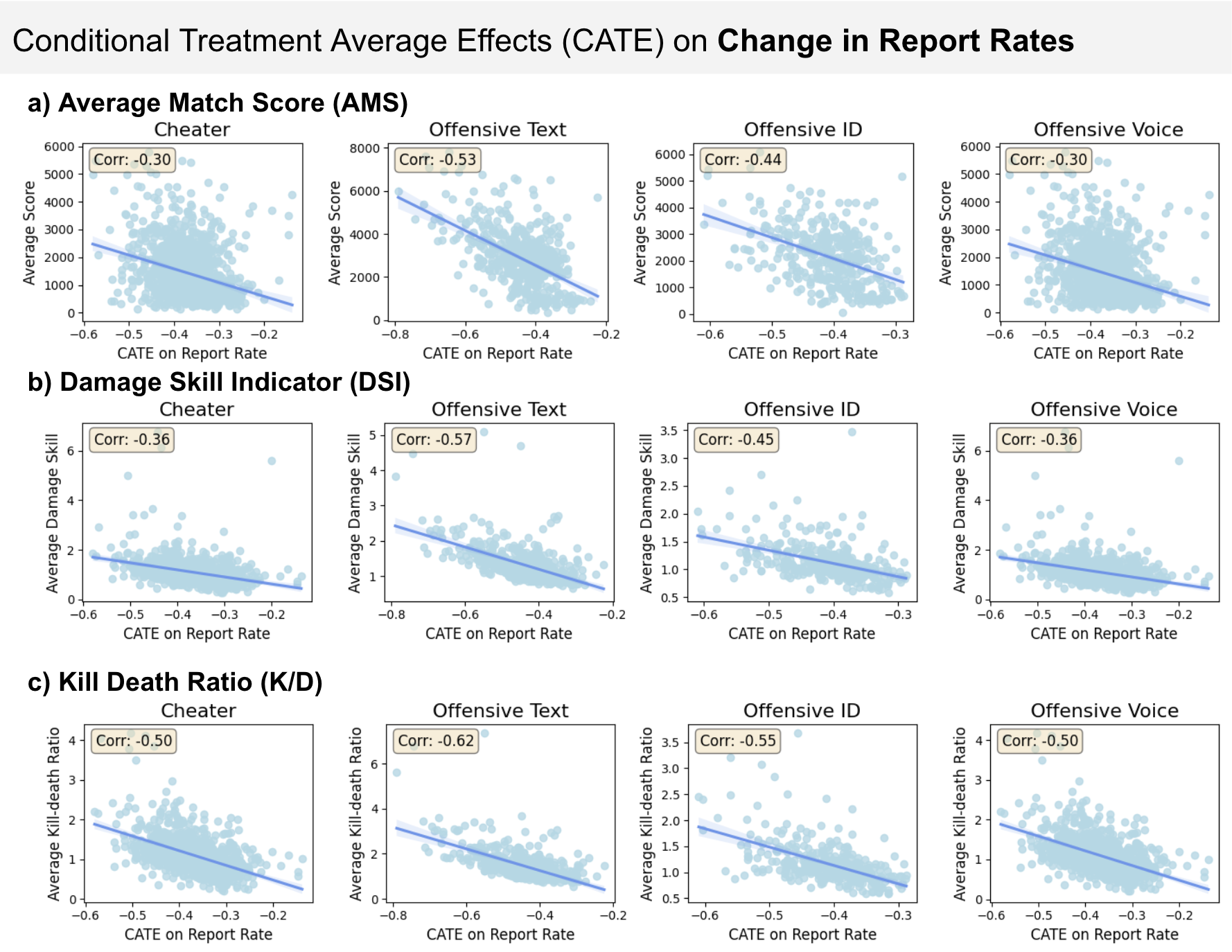}

    \centering

\end{figure}

\begin{figure}[htbp]
    \caption{Correlations of conditional average treatment effect (CATE) for change in Participation Rates and player performance characteristics.}
    \Description{Correlation of CATE for Report Rates and Player Performance Characteristics.}
    \label{fig:cate_on_participation}
    \includegraphics[width=.80\linewidth]{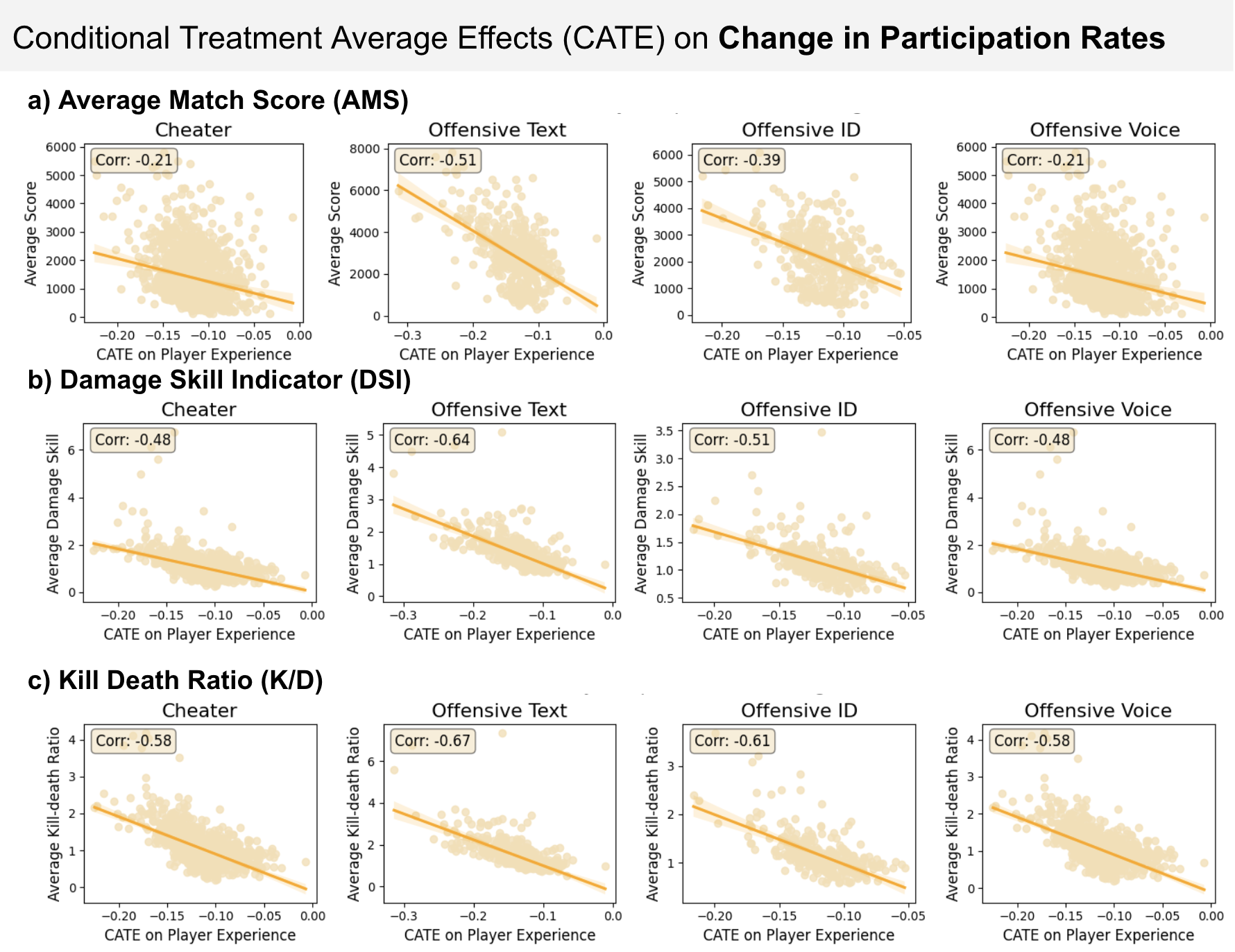}

    \centering

\end{figure}

\end{document}